\documentclass{article}
\usepackage[letterpaper,top=2cm,bottom=2cm,left=3cm,right=3cm,marginparwidth=1.75cm]{geometry}
\usepackage{amsmath}
\usepackage{amsthm}
\usepackage[defaultsups]{newtxtext}
\usepackage[varg]{newtxmath}
\usepackage{bm}
\usepackage{graphicx, xcolor} 
\usepackage{subcaption}
\usepackage{here}
\usepackage{natbib}
\usepackage[colorlinks=true, allcolors=blue]{hyperref}

\begin{document}
\title{Distance between Road Networks: A Macroscopic Method for Road Network Datasets Comparison Using Traffic-weighted Geographic Distribution}
\author{
  Hengyi Zhong$^1$\\
  \texttt{zhong.h.aa@m.titech.ac.jp}
  \and
  Toru Seo$^1$\\
  \texttt{seo.t.aa@m.titech.ac.jp}
}
\date{$^1$Institute of Science Tokyo}
\maketitle

\begin{abstract}
In transportation network analysis, various types of road network data can be used even when focusing on the same region.
Since different road network datasets can make different performance in analyses, it is necessary to compare them and make appropriate selections in a qualitative manner.
However, many of the existing methods for comparing road network datasets are limited to specific topological evaluations and do not consider transportation.
This study proposes a method for quantitative comparison of different road network datasets with explicit consideration for traffic flows on them.
The method first conducts a static traffic assignment with hypothetical demand for each dataset, and then compare the results using Wasserstein distance on two dimensional plane.
Case study on different sources of road network datasets and their simplifications suggests the potential use of the proposed method in evaluating and selecting road network datasets.
\end{abstract}


\section{Introduction}
In transportation network analysis, road network data plays an important role as the foundation where all travelers move and traffic flows.
Road network data is a type of geographic information system (GIS) data which organizes real-world roads into a network structure with their connections and shapes.
Nowadays, an increased accessibility to various types and formats of road network data, such as OpenStreetMap (OSM) to which generated from Global Navigation Satellite System (GNSS) trajectory data \citep{Huang_2018, wang_novel_2015} and satellite imageries \citep{Hinz_2003}, has dramatically improved the efficiency of research on road networks, including transportation network analysis.

\begin{figure}
  \begin{center}
    \includegraphics[width=10cm]{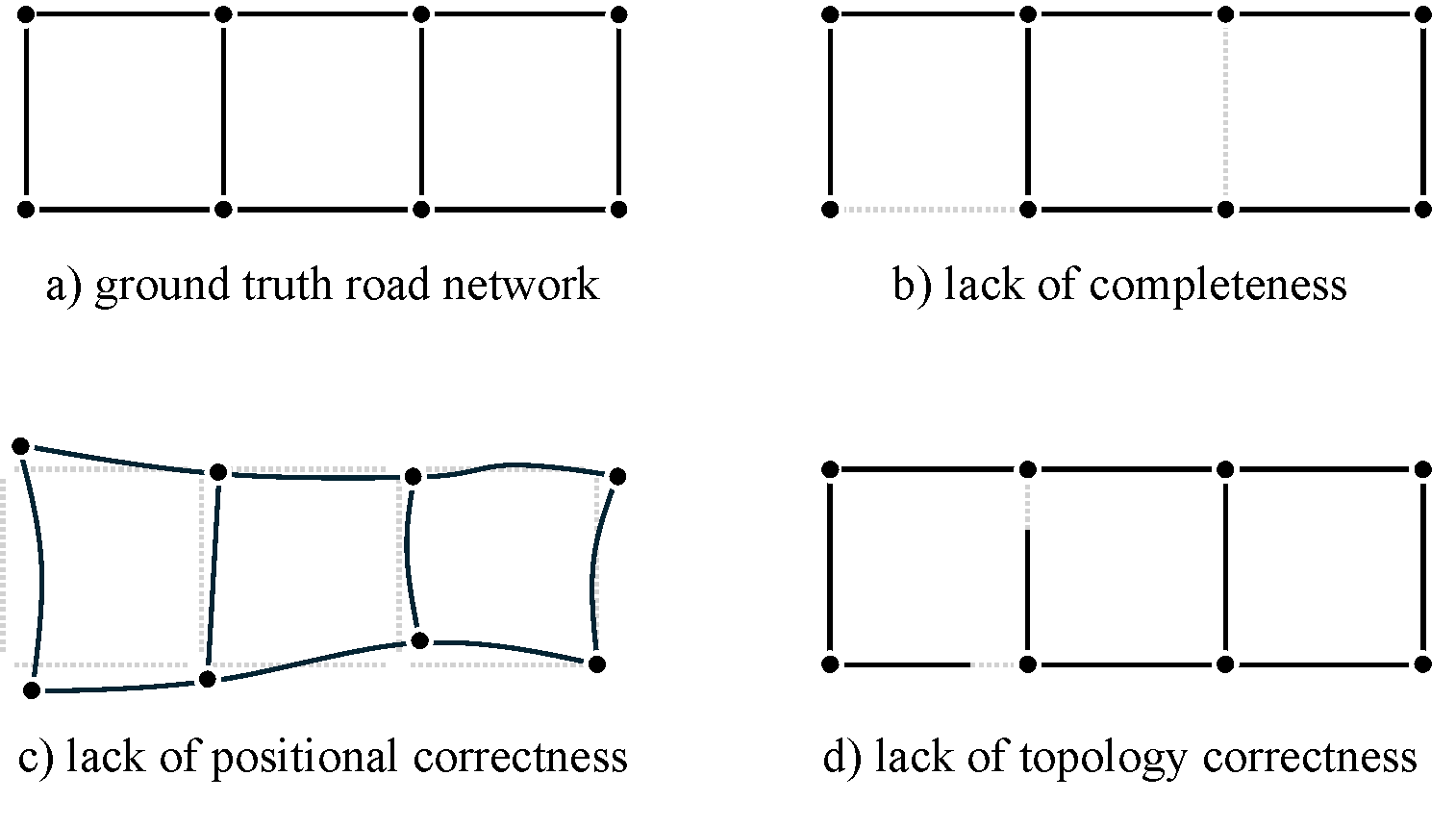}\\
 \vskip\baselineskip
  \end{center}
  \caption{Common problems in road network datasets}
 \label{link_lack}
\end{figure}

The comparison of different road network datasets is essential for understanding the quality of data and evaluating their accuracy and usefulness \citep{Oort_2006}.
For the comparison of road network datasets, previous studies have argued that attention should be given to the quality aspects of network, such as positional accuracy, completeness and topological correctness \citep{Haklay_2010, Hashemi_2017}.
These metrics are essential for describing common problems on road network datasets shown in Figure \ref{link_lack}, which measure how much a road network dataset could represent the corresponding ground-truth road network geographically and topologically.

To make comparison between two road network datasets, matching processes of corresponding links and nodes between them are required in these methods.
Previous studies expected that datasets for comparison have almost the same components refer to the ground truth, so that most of the links and nodes in one dataset have their corresponding ones in the comparison target dataset \citep{Haklay_2010}.
However, matching processes can be difficult in analysis using different kinds of road network datasets.
For large-size road network datasets such as detailed regional road network, matching of each link and node has large time and computational cost.
Meanwhile for those aggregated and simplified road networks usually used in transportation analysis, it is hard to match each component as a large part of links and nodes were merged or omitted to make a smaller size of road network data.

\begin{figure}
  \begin{center}
    \includegraphics[width=10cm]{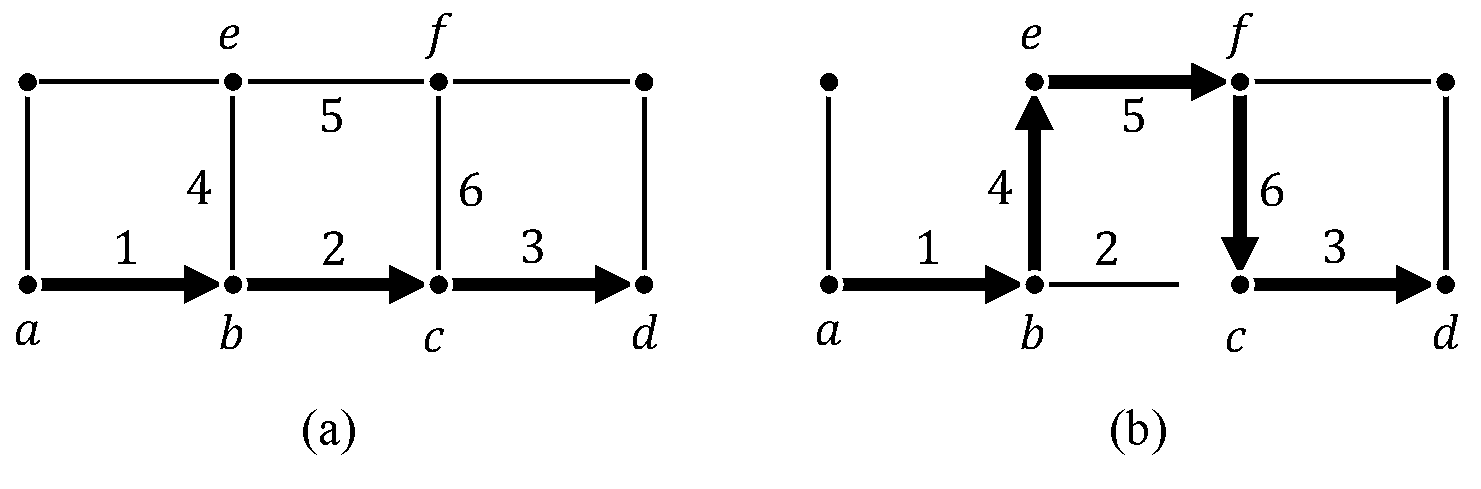}\\
 \vskip\baselineskip
  \end{center}
  \caption{Effect on traffic states caused by low topological correctness}
 \label{topo}
\end{figure}

Furthermore, these methods for comparing road network datasets in previous studies only focus on a specific feature of network and may not be appropriate to directly apply to transportation network analysis, in which various local and global features are essential.
In transportation analysis, features for comparison can be varied.
But in general, instead of performances of each link and node, overall distribution of traffic states on the road network is more important for the evaluation of road network datasets for those analysis purposes.
For example, lack of topological correctness causes detours on the shortest path (Figure \ref{topo}), which changes the distribution of overall traffic state.
Meanwhile topologically correct simplified road networks may not accurate, but tends to computationally efficient and are expected to keep the original traffic status.
In this case, existing methods could only evaluate the accuracy but not traffic and computational issues, thus the usefulness of data could not be evaluated properly.

In this paper, we developed a method of comparing road network datasets at a macroscopic scale.
Here, \textit{macroscopic} refers to the perspective of the comparison method: rather than evaluating individual links and nodes, the method evaluates the network as a whole through the spatial distribution of its traffic states.
This is in contrast to \textit{element-level} methods that compare matched links and nodes one by one.
The proposed method catches one road network dataset including both its geographic and traffic attributes as a whole, so that it can capture the global effects of aspects for transportation network analysis.
Traffic distribution on a road network is considered as a geographic distribution weighted by traffic parameters.
Then differences between two road networks is measured by a traffic-weighted distance between two corresponding distributions.
Also, a possible application on evaluating different road network data in network extraction by the proposed method was discussed.

The contribution of this paper can be summarized as follows: 
\begin{enumerate}
\item A new framework for comparison of road network datasets at a macroscopic scale is proposed.
\item To enable quantitative comparison, we propose a novel metric, termed {\it TG (traffic-weighted geographic) distribution}.
The method transforms quality aspects of road network datasets into their effects on spatial traffic distributions, capturing global transportation characteristics emerging from element-level network structures.
\item Based on above formulation, we further define the {\it TGW (traffic-weighted geographic Wasserstein) distance} for measuring differences between datasets and the {\it TG-OTM (traffic-weighted geographic optimal transport matrix)} for detecting their spatial discrepancies.
\item The proposed method enables comparison of road network datasets for the same area without explicit matching of network elements.
It is applicable to simplified or aggregated datasets where corresponding components are difficult to identify.
\end{enumerate}

Note that the proposed method does not aim to substitute the existing methods such as completeness.
Rather, it aims to complement them.
For example, in comparison among simplified road networks, existing methods evaluates where and how much it was simplified compared to the original one.
Additionally, the proposed method evaluates how overall traffic aspects change by those element-level aspects.
By using the proposed method and the existing methods simultaneously, it is helpful to understand both local and global features of the dataset.

This paper is organized as follows. 
Section 2 reviews previous studies in comparison of road network datasets and works relates to the distance between distributions.
Section 3 explains the proposed methodology to compare road network datasets at a macroscopic scale, and describes the characteristics of comparisons by traffic distributions.
Section 4 shows case studies of comparisons between different road network datasets used in transportation network analysis, followed by conclusions in section 5.

\section{Literature reviews}
In this section, reviews on existing comparison methods of road network datasets are summarized.

\subsection{Different road network datasets}
Nowadays, a large number of road network datasets provided by different resources are available.
Traditionally, road network datasets are generated by the digitalization existing road network maps, which are constructed mainly by surveying.
For instance, the Digital Road Map (DRM), which is widely used as a standard in Japan, is based on topographic maps from the Geospatial Information Authority of Japan, with using blueprints of roads provided by road administrators \citep{Kanasugi_2019}.

With the development of Web 2.0 in the early 21st century, services providing geographic information including road network datasets via the Internet have greatly advanced.
A wide variety of map services have become available, ranging from geospatial information provided by public institutions to commercial maps such as Google Maps and TomTom \citep{Hakley_2008}.
In recent years, OpenStreetMap (OSM), created by the collaborative efforts of volunteers worldwide, has attracted significant attention and is widely utilized due to its ease of access and use as an open-source dataset \citep{Haklay_2010, Toya_2010, Kanasugi_2019}.

Meanwhile, road networks extracted from imagery and traffic data are also popular among GIS researchers.
Since those data can be collected quickly with high accuracy, such road network generation methods can be efficient solutions for obtaining data of road networks.
Satellite imagery is essential on extracting up-to-date road networks from road centerlines \citep{Hinz_2003, Wang_2016r, Miao_2014, Gao_2019} to sidewalks \citep{Ning_2022}.
Traffic trajectory-based road network generation can be utilized to generate navigable road network data \citep{Huang_2018, wang_novel_2015}.
Those datasets can be in different scales, varies from detailed lane level networks \citep{uduwaragoda_generating_2013} to aggregated networks \citep{Zhong_2023}.

In using such a wide range of road network datasets, how to evaluate and select the appropriate dataset which not only represents the ground-truth road network well but also has good computational performance remains as a problem.

\subsection{Comparison methods of road network datasets}
Comparison of road network datasets is important to understanding the quality of dataset as geographic information \citep{Oort_2006}.

Previous discussions on comparing and evaluating road network datasets mainly focusing on the characteristics of network, such as positional accuracy, completeness and topological correctness \citep{Haklay_2010, Hashemi_2017}.
\cite{Haklay_2010} evaluates the goodness of volunteered geographic information, and proposed completeness matrices regarding to positional accuracy and total length of roads in road network datasets.
Quality of OpenStreetMap (OSM), a popular volunteered geographic information, is evaluated by comparing to Meridian dataset, a survey-based road network dataset in United Kingdom.
Furthermore, \cite{Toya_2010} evaluates the temporal correctness of OSM by considering the edit history on time-series changes of links and nodes.

Similarly, comparisons for data-oriented generated road networks are important to evaluate the usefulness of their generation methods.
For evaluating data-oriented generated networks from imaginary or traffic data, \cite{Hashemi_2017} reviewed and proposed a generic testbed focusing on differences of those road networks to the ground truth network, instead of simply overlaying on the ground truth.
It has been used in evaluations of several trajectory-based road networks \citep{guo_novel_2022, zhou_pedestrian_2021}.
In this testbed, positional accuracy is defined as the average geological distance between corresponding links and nodes, and completeness is defined as the matching rate of them.
Topological correctness is defined as the average number of links along the shortest path, which evaluates the connectivity of links and nodes.

However, those element-level evaluation methods only focus on a specific feature of network, which could not be used in understanding the overall performance of the network.

png\section{Methodology}
\subsection{Overview}

\begin{figure}
  \begin{center}
    \includegraphics[width=16cm]{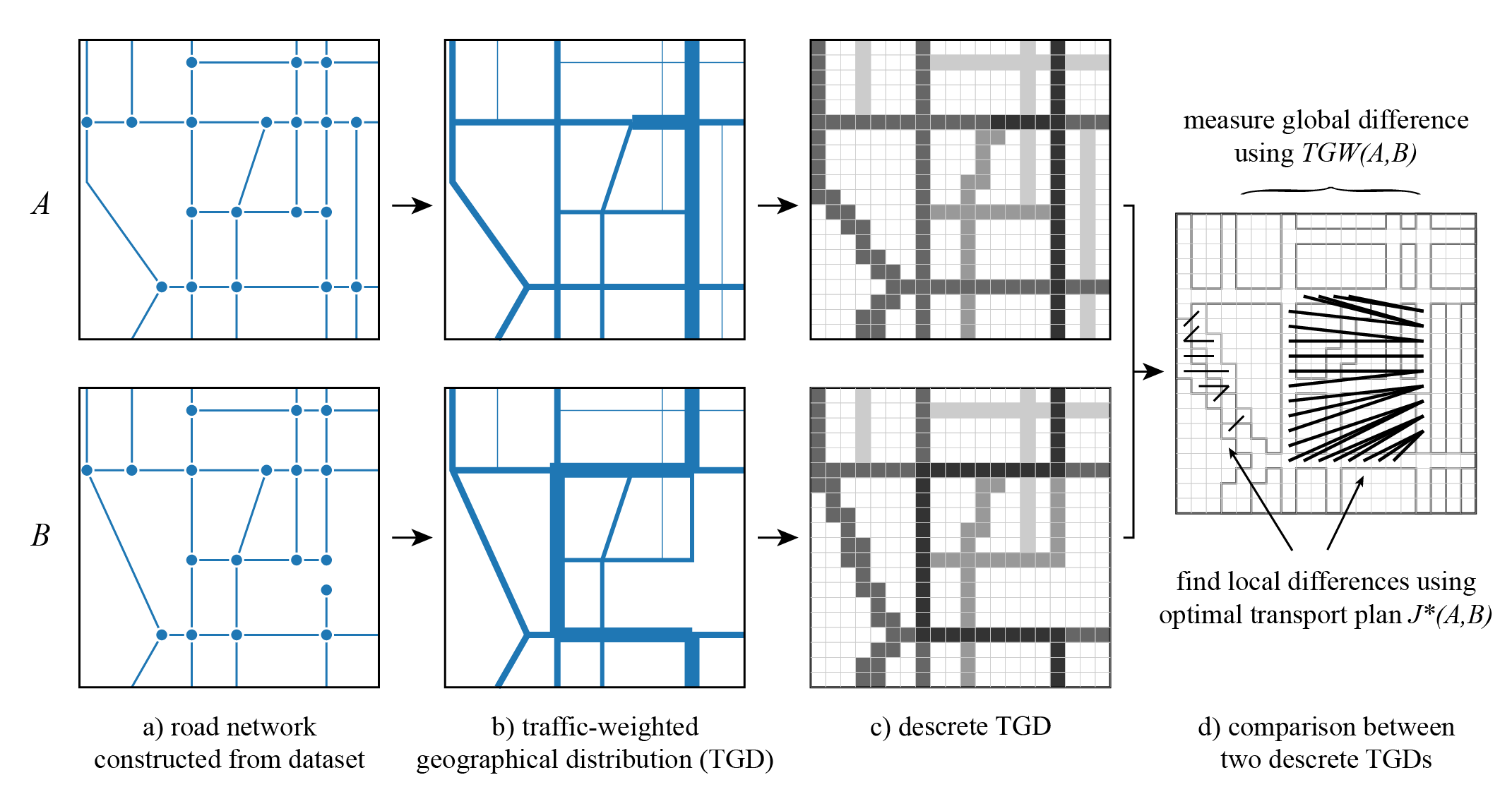}\\
 \vskip\baselineskip
  \end{center}
  \caption{Overview of road network dataset processing in the proposed comparison method}
 \label{flow}
\end{figure}

\begin{figure}
  \begin{center}
    \includegraphics[width=16cm]{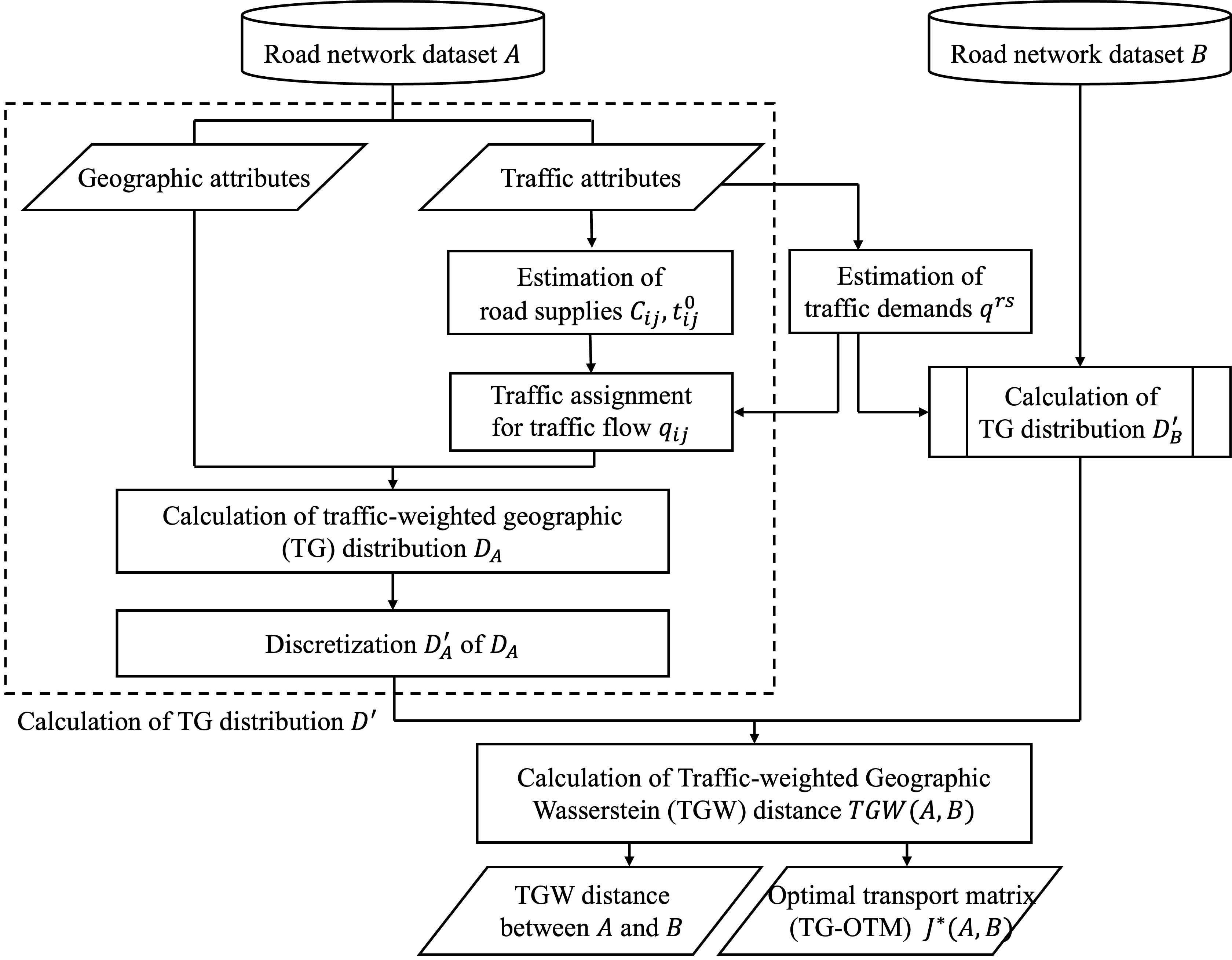}\\
 \vskip\baselineskip
  \end{center}
  \caption{Flowchart of proposed method for comparing datasets $A$ and $B$}
 \label{chart}
\end{figure}

The proposed method evaluates road network dataset from geographic distribution of the road network it represents (Figure \ref{flow}).
To evaluate a road network dataset at a macroscopic scale, one road network dataset is converted to only one geographic distribution.
Differences between two datasets can be measured by the similarity between two corresponding geographic distributions.
Flowchart of proposed method for comparing two road network datasets $A$ and $B$ is shown in Figure \ref{chart}.

Not all links perform the same role in transportation network.
Therefore, the aforementioned geographic distribution should be weighted appropriately according to their importance.
To quantify the link importance, we propose the use of traffic volume of each link, which can be predicted by using traffic engineering methodology called traffic assignment \citep{Sheffi_1984}.

By using traffic volume as the weight for link importance (Figure \ref{flow}(b)), the proposed method evaluates the change of traffic by the differences of all assigned routes.
Consider a travel from one origin to one destination on the shortest path between them.
The route of that path can be different on two road network datasets due to differences on positional accuracy, topology and traffic parameters.
This difference is measured by the area between two routes and the traffic volume of that path.
The proposed method accumulates differences of shortest paths for all possible pairs of origins and destinations, in a comprehensive way by assigning and evaluating overall traffic across the road network.

Traffic assignment method is applied to obtain possible link traffic volumes on road network datasets.
Parameters of supplies and demands is necessary for assigning traffic on road network, and can be derived solely from the road network dataset appropriately.
Supplies include traffic characteristic values of links, such capacity and free flow travel time. Those values can be set by the road type for each link.
Demands mean how much traffic are made for each pair of origin and destination, which can be estimated from the structure of road network.

\begin{figure}
  \begin{center}
    \includegraphics[width=10cm]{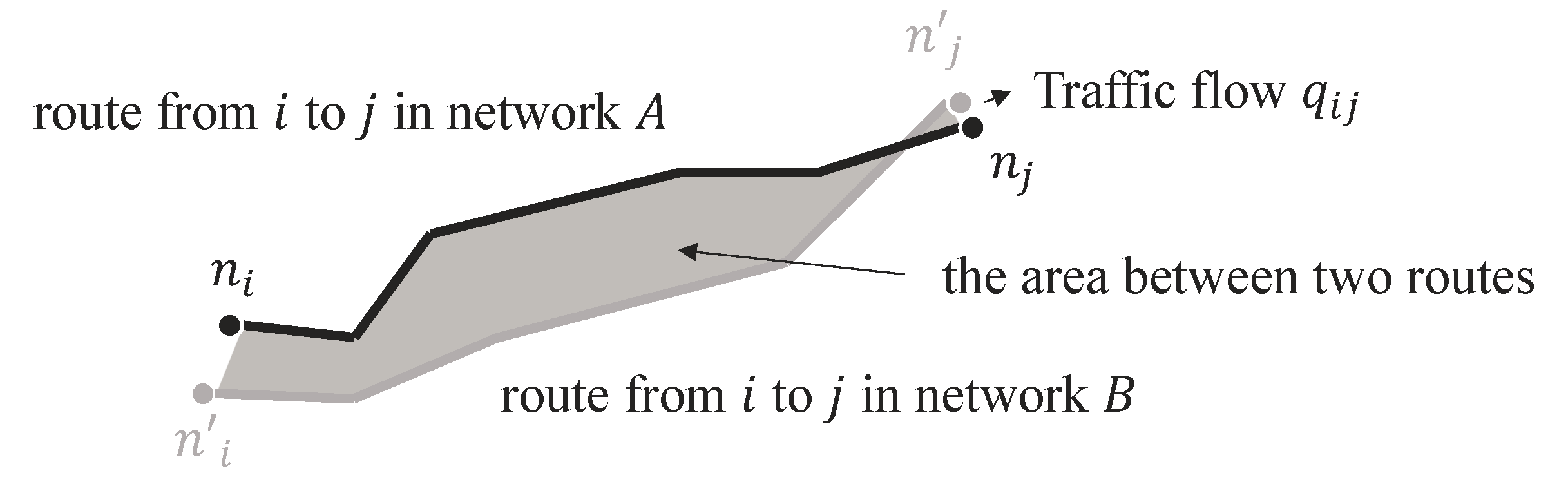}\\
 \vskip\baselineskip
  \end{center}
  \caption{Factors on evaluating similarity of traffic states distribution between two datasets}
 \label{fce}
\end{figure}

Based on estimated traffic states, we construct a novel spatial distribution that represents the traffic states of a road network dataset along its geometry, termed the \textit{TG (traffic-weighted geographic) distribution} (Figure \ref{flow}(b,c)).
In what follows, the prefix \textit{TG-} indicates objects defined on this traffic-weighted geographic representation.
The distance between two TG distributions (Figure \ref{flow}(d)) is measured by the Wasserstein distance, which we call the \textit{TGW (traffic-weighted geographic Wasserstein) distance} between the two road network datasets.
Wasserstein distance is a distance function evaluates the similarity between two distributions, which catches the shape of distributions well \citep{Ambrosio_2009}.
The similarity is evaluated by traffic flows of routes and the area between their corresponding positions in two networks (Figure \ref{fce}), over the entire study area.
A small value of TGW distance can be expected if the TG distributions of two datasets are similar.
Furthermore, the optimal transport matrix $J^*(A,B)$ that attains this TGW distance, the \textit{TG-OTM (traffic-weighted geographic optimal transport matrix)}, is visualized to localize spatial discrepancies between the two networks.

\subsection{Formulation}
\subsubsection{Road network dataset}
In this study, a road network dataset $\mathcal{D}$ is represented as a graph consisting of a set of nodes $\mathcal{N} = \{n_i\}$ as multipoint and a set of directed links $\mathcal{E} = \{e_{ij}\}$ as polyline.

\begin{figure}
  \begin{center}
    \includegraphics[width=5cm]{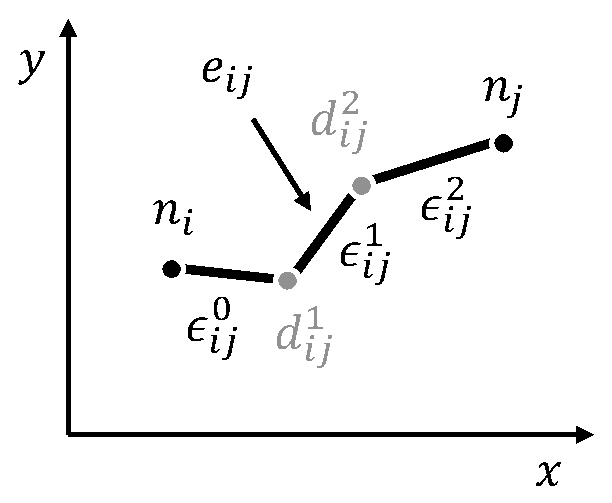}\\
 \vskip\baselineskip
  \end{center}
  \caption{Link $e_{ij}$ in geographic coordinates}
 \label{fig_link}
\end{figure}

Each node $n_i$, with coordinates $(x(n_i), y(n_i))$, represents an intersection.
Each node is connected to at least one link and serves as the start or end point of a link.

A directed link $e_{ij}$ (Fig. \ref{fig_link}) represents a road segment from node $n_i$ to node $n_j$.
The shape of the link is described by a sequence of interior vertices $d_{ij}^1, ..., d_{ij}^k$, each with coordinates $(x(d_{ij}^a), y(d_{ij}^a))$.
These vertices divide the link into straight segments $\epsilon_{ij}^a = \epsilon(d_{ij}^a, d_{ij}^{a+1})$.

Each link has attributes such as length, capacity, and speed limit.
Traffic-related values, such as traffic volume, travel time, or speed, are assigned to each link as a traffic weight $w_{ij}$.

\subsubsection{Traffic assignment}\label{ta}
Evaluating the distribution of traffic states on a road network dataset is helpful for understanding its performance in representing real-world road traffic in transportation analysis.
Quality aspects of the dataset affect how traffic states are distributed, leading to differences in estimated distributions among different road network datasets.

Estimation of traffic state distributions is performed by traffic assignment methods.
A basic framework of traffic assignment, static traffic assignment with user equilibrium \citep{Sheffi_1984}, is applied in this study.
Given traffic demands between all origin-destination (OD) pairs within the road network, it estimates traffic flows across the entire network.
The estimated flows satisfy an equilibrium in which all travelers choose routes that minimize their own travel time, considering congestion effects.
Traffic assignment is applied to each network separately, determining traffic flow $q_{ij}$ as the weight $w_{ij}$ of each link $e_{ij}$ for each dataset.

Static traffic assignment is based on Wardrop's first principle \citep{Wardrop_1952}.
It assumes that each driver chooses the route with the minimum travel time from origin to destination.
This leads to the result that, for all routes between a given OD pair, travel times of all used routes are equal and are not longer than those of unused routes.
In this case, no traveler can reduce their travel time by changing routes, and an equilibrium is reached.
Given OD demands, static traffic assignment calculates link traffic that satisfies Wardrop's first principle over the network.

To reproduce congestion on roads, a link performance function is used to define link travel time $t_{ij}$ under a given traffic volume $q_{ij}$ and capacity $C_{ij}$.
The Bureau of Public Roads (BPR) function \citep{us1964traffic}, a widely used power function, is adopted as the link performance function.
\begin{align}
\label{bpr}
  t_{ij}(q_{ij})= t_{ij}^0\{1+\alpha(\frac{q_{ij}}{C_{ij}})^\beta \}
\end{align}
where $t_{ij}^0$ is the free-flow travel time when $q_{ij}=0$.
$\alpha$ and $\beta$ are empirical parameters and can be set based on common practice \citep{BRANSTON1976, Spiess_1990}.

Traffic flows under user equilibrium are obtained by minimizing the total link travel time in the following equivalent convex optimization problem.
\begin{align}
  \min \quad & Z(q) = \sum_{ij} \int_{0}^{q_{ij}} t_{ij}(q)dq \\
  \textrm{s.t.} \quad & \sum_{ij} q_{ij}^{rs} = q^{rs} \\
  & q^{rs} \leq 0
\end{align}
where $q^{rs}$ is the OD demand between OD pair $r$ and $s$, and $q_{ij}^{rs}$ is the traffic on link $e_{ij}$ associated with routes between OD pair $r$ and $s$.
The Frank-Wolfe algorithm is used to solve the static traffic assignment problem \citep{Sheffi_1984}.

This traffic assignment approach provides stable estimates of traffic flows and is useful for evaluating road network performance \citep{Sheffi_1984}.
In static traffic assignment, traffic states are reproduced under time-independent assumptions of supply and demand, while allowing application to relatively large networks \citep{Leblanc_1975}.

\subsubsection{Supplies and demands for traffic assignment}\label{tb}

Necessary traffic characteristics (e.g., capacity and limit speed) and OD demands for traffic assignment in Section~\ref{ta} can be hypothetically derived from the road network data.

Traffic characteristics are assigned to each link according to its road type.
Links of higher-class road types, such as trunk roads and motorways, are assigned larger capacity and higher speed limits, and vice versa.

OD demand is estimated from assumed trip generation and attraction volumes within the study area using a gravity model \citep{voorhees_1955}.
The study area is divided into 1km $\times$ 1km zones, and each zone $z$ is assigned trip attraction volume $A_z$ and trip generation volume $G_z$.
A centroid is defined for each zone, and the shortest travel times $c_{rs}$ between zones are calculated.
Node $n_z$ representing the centroid of zone $z$ is defined as the node with the smallest geographic distance to the centroid.
Traffic volume $q^{rs}$ from zone $r$ to zone $s$ is determined by the gravity model:
\begin{eqnarray}
\label{gravity}
q^{rs} = kG_r^{\alpha}A_s^{\beta}c_{rs}^{\gamma}
\end{eqnarray}

To represent a large number of external trips (e.g., those passing through expressways) when the study area is part of a megacity, higher trip generation and attraction volumes are assigned to zones connected to major external roads, and lower volumes to other zones.

When comparing two or more road network datasets, the same OD demand should be used to ensure fair comparison under identical demand conditions.
This ensures that differences in traffic states arise solely from network structure.
Such a pattern can be derived from the dataset that is most likely to be closest to the ground truth.

\subsubsection{Traffic-weighted geographic distribution}

We convert the traffic assignment results with their geographic attributes into a two-dimensional distribution, termed {\it TG (traffic-weighted geographic) distribution}.
Specifically, the TG distribution of a road network dataset represents spatial traffic states distributed along the geometry of each link.
Geographical attributes of links are contained in the dataset, and traffic states are obtained by traffic assignment in sections \ref{ta} and \ref{tb}.

The geographic distribution $D$ of a road network dataset $\mathcal{D}$ is defined as a set of distributions $E_{ij}$ for each link $e_{ij}$.
$D$ is defined over a two-dimensional space of size $M \times N$, corresponding to the study area $\mathcal{R}: x \in (x_0, x_0+M), y \in (y_0, y_0+N)$.

Each link $e_{ij}$ is distributed along its geometry, shown as the black path in Figure \ref{fig_link}.
It consists of straight segments $\epsilon_{ij}^a$ divided by interior vertices.
Each segment $\epsilon_{ij}^a$ lies on the straight line between two interior vertices $d_{ij}^a$ and $d_{ij}^{a+1}$.
The corresponding distribution $E_{ij}^a$ of segment $\epsilon_{ij}^a$ is described by the two-point form of a line in geographic coordinates $(x,y)$:
\begin{align}
  \frac{x-x(d_{ij}^a)}{x(d_{ij}^{a+1})-x(d_{ij}^a)} = \frac{y-y(d_{ij}^a)}{y(d_{ij}^{a+1})-y(d_{ij}^a)}, x \in [x(d_{ij}^a),x(d_{ij}^{a+1})]
\end{align}

Since $E_{ij}$ is the set of all segment distributions $E_{ij}^a$, $D$ can be represented by these segment-level distributions.

Assuming that traffic exists only on road segments, traffic weight $w_{ij}$ on $\mathcal{D}$ is distributed only on the corresponding $E_{ij}$.
Thus, for any point $(x,y) \in \mathcal{R}$, if $(x,y)$ lies on link $e_{ij}$, it takes the traffic weight $w_{ij}$ of that link.
The TG distribution $D$ of dataset $\mathcal{D}$ is defined as:
\begin{align}
\label{road_dist}
  D: (x,y) = w_{ij},  \forall (x,y) \in E_{ij}, \forall e_{ij} \in \mathcal{E}
\end{align}

\subsubsection{Traffic-weighted geographic Wasserstein distance}
The distance between two networks' TG distributions is measured as {\it TGW (traffic-weighted geographic Wasserstein) distance}.
The TGW distance $TGW(A,B)$ between two road network datasets $\mathcal{D}_A$ and $\mathcal{D}_B$ is defined as the Wasserstein distance $W(D_A,D_B)$ between their corresponding TG distributions $D_A$ and $D_B$.

The Wasserstein distance is a metric that evaluates the similarity between two distributions $A$ and $B$ while considering their geometric structure \citep{Cuturi_2015, Ambrosio_2009, Bonneel_2015}.
It satisfies the axioms of a distance and captures spatial characteristics of distributions.
In general, when two distributions have similar shapes and are located close to each other, the Wasserstein distance takes a small value.

The Wasserstein distance $W(A,B)$ can be interpreted as the minimum transport cost required to transform $A(x)$ into $B(y)$ via a transport matrix $J(A,B)$:
\begin{align}
  W(A,B) = \inf_{J \in \mathcal{J}(A,B)} \int C(x,y)dJ(x,y)
\end{align}
where $J(A,B)$ is a joint distribution of $A$ and $B$, and $\mathcal{J}(A,B)$ is the set of all such transport matrices.
$C(x,y)$ is a cost function representing the cost of moving one unit of weight from $x$ to $y$.
In this study, the weight corresponds to the traffic weight $w_{ij}$, and the cost function is the Euclidean distance between positions $x$ and $y$.

In the standard Wasserstein distance, total weights $\int A dx$ and $\int B dy$ must be equal so that there exists $J(A,B)$ satisfying the marginal constraints $\int J(x,y)dy = A$ and $\int J(x,y)dx = B$.
However, total weights of TG distributions are generally not equal across different road network datasets, even under the same OD demand.
The total weight of a TG distribution is proportional to the total travel distance over the network, and it increases when routes are detoured and decreases when links are unused.

To relax this constraint, unbalanced Wasserstein distance \citep{chizat2016} is used to define the TGW distance.
A penalty for mass variation is introduced in the transport cost:
\begin{align}
\label{was_dist}
  W(A,B) = \inf_{J \in \mathcal{J}(A,B)} \int C(x,y)dJ(x,y) + \lambda (D(\int J(x,y)dy, A) + D(\int J(x,y)dx, B))
\end{align}
where $\lambda$ controls the penalty for mass variation.
The divergence term $D$ measures the discrepancy between transported distributions $\int J(x,y)dy, \int J(x,y)dx$ and the original distributions $A, B$, respectively.
The Kullback--Leibler divergence $KL(\int J(x,y)dy || A) + KL(\int J(x,y)dx || B)$ is commonly used for $D$ \citep{chizat2019}.

\subsubsection{Traffic-weighted geographic optimal transport matrix}
To detect regions where distribution of traffic states have large differences between TG distributions, the optimal transport matrix corresponding to the TGW distance is analyzed and visualized.

In Equation \ref{was_dist}, $J(x,y)$ represents the transport matrix describing the amount of mass transported from the TG distribution $D_A$ to $D_B$, and $\mathcal{J}(D_A,D_B)$ denotes the set of all such transport plans.
Among them, the optimal transport plan $J^*(D_A,D_B)$ minimizes the total transport cost.
This optimal transport matrix $J^*(D_A,D_B)$ is referred to as the {\it TG-OTM (traffic-weighted geographic optimal transport matrix)}, and its transport cost is equal to the TGW distance.

The TG-OTM describes how the traffic weight at each location in $D_A$ is redistributed to locations in $D_B$.
By visualizing the TG-OTM in the geographic space of the road network, regions with both large transport distance and large transported mass can be identified.
This enables the detection of local differences between road networks.

\subsubsection{Numerical computation method for TGW distance}
To numerically compute the TGW distance and the corresponding TG-OTM $J^*(A,B)$, the study area is discretized into grids so that the TG distribution $D$ can be represented as a two-dimensional matrix.

A road network in a modern city or region typically contains thousands of links, which would require a large number of equations in the continuous formulation (Equation \ref{road_dist}).
Computing the Wasserstein distance for such a system is computationally expensive.
Instead, the discretized matrix representation is more suitable for numerical computation.

For the continuous study area $\mathcal{R}: x \in (x_0, x_0+M), y \in (y_0, y_0+N)$, square grids of size $C \times C$ are used to obtain a discrete region $\mathcal{R}'$ with $m \times n$ cells, where $m=M/C$, $n=N/C$, and $m,n \in \mathbb{N}$.
Using $\mathcal{R}'$, the discrete distribution $E_{ij}'$ of link $e_{ij}$ is defined as the set of grid cells $(p,q)$ through which the corresponding line segments pass.
For each point $(x,y) \in \mathcal{R}$, the corresponding grid $(p,q) \in \mathcal{R}'$ is given by:
\begin{align}
  p &= \frac{x-x_0}{C} \\
  q &= \frac{y-y_0}{C}
\end{align}

\begin{figure}
    \begin{minipage}{.50\textwidth}
        \centering
        \includegraphics[width=3.5cm]{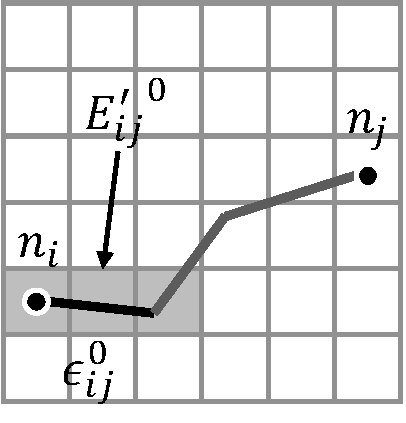}
        \caption{Discretization of link segment $\epsilon_{ij}$}
        \label{f1a}
    \end{minipage}
    \def\@captype{table}
    \begin{minipage}{.50\textwidth}
        \centering
        \includegraphics[width=3.5cm]{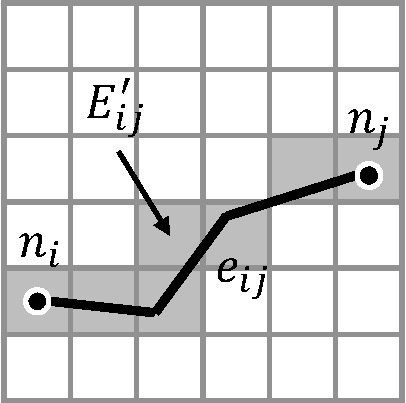}
        \caption{Discretization of link $e_{ij}$}
        \label{f1b}
    \end{minipage}
  \end{figure}

The TG distribution is then represented as a grid-based discrete distribution.
For each segment $\epsilon_{ij}^a$ of link $e_{ij}$, the grids corresponding to its interior vertices $d_{ij}^a$ and $d_{ij}^{a+1}$ are identified.
The grid cells along the straight line between these two grids correspond to the segment $\epsilon_{ij}^a$ (Figure \ref{f1a}), and are defined as $E_{ij}^{'a}$.
The distribution $E_{ij}'$ of link $e_{ij}$ is given by the union of all segment distributions (Figure \ref{f1b}):
\begin{align}
  E_{ij}' = E_{ij}^{'0} \cup E_{ij}^{'1} \cup ... \cup E_{ij}^{'k}
\end{align}

As in the continuous case, traffic weight $w_{ij}$ is assigned to the corresponding $E_{ij}'$.
Thus, the discrete TG distribution $D'$ of dataset $\mathcal{D}$ is defined as:
\begin{align}
\label{disc_dist}
  D': (p,q) = w_{ij},  \forall (p,q) \in E_{ij}', \forall e_{ij} \in \mathcal{E}
\end{align}

This discrete TG distribution $D'$ can be represented as an $m \times n$ matrix.
Using discrete TG distributions $D_A'(\boldsymbol{x})$ and $D_B'(\boldsymbol{y})$, the TGW distance in Equation \ref{was_dist} and the corresponding optimal transport plan $J^*(A,B)$ can be computed as:
$D_A'(\boldsymbol{x})$ and $D_B'(\boldsymbol{y})$.
\begin{align}
\label{traffic_was}
  TGW(A,B) = W(D'_A(\boldsymbol{x}),D'_B)(\boldsymbol{y})
\end{align}
where $\boldsymbol{x}=(p_A, q_A)$ and $\boldsymbol{y}=(p_B, q_B)$ denote grid indices in $D_A'$ and $D_B'$, respectively.

\subsubsection{Summary}
The proposed method compares two road network datasets $\mathcal{D}_A$ and $\mathcal{D}_B$ by following steps:
\begin{enumerate}
\item Calculate OD demands for the area of two road network datasets.

\item Estimate traffic states on $\mathcal{D}_A$ and $\mathcal{D}_B$ by traffic assignment to obtain the weight of traffic $w_{ij}$ for each link, respectively.

\item Generate discrete TG distribution $D_A', D_B'$ of $\mathcal{D}_A, \mathcal{D}_B$, respectively.

\item Calculate the TGW distance $TGW(A,B)$ with its TG-OTM.

\item Evaluate the similarity of $\mathcal{D}_A$ and $\mathcal{D}_B$ by the value of $TGW(A,B)$ and the distribution of TG-OTM.
\end{enumerate}

png\subsection{Features, advantages, and disadvantages of the proposed method}

Features of proposed method are shown by discussing the meaning of the TGW distance in using traffic-weighted geographic distributions.
Then the relationships of proposed method to existing metrics are discussed in three major quality aspects for comparisons, which are completeness, positional accuracy and topological correctness.
This describes how proposed method evaluates road network datasets at a macroscopic scale.

\subsubsection{Meaning of the distance}
In this section, the features of the proposed method are discussed by focusing on the unit and interpretation of the TGW distance between TG distributions.

The Wasserstein distance $W(A,B)$ is defined as the minimum cost required to transform distribution $A$ into $B$, computed as the integral of the product of a weight term and a distance term.
As defined in equation (\ref{was_dist}), this cost is obtained by integrating the transported mass $dJ(x,y)$ and the transported distance $||x-y||$.
Therefore, the dimension of the Wasserstein distance is given by the product of the dimensions of the weight term and the distance term.

\begin{figure}
  \begin{center}
    \includegraphics[width=6cm]{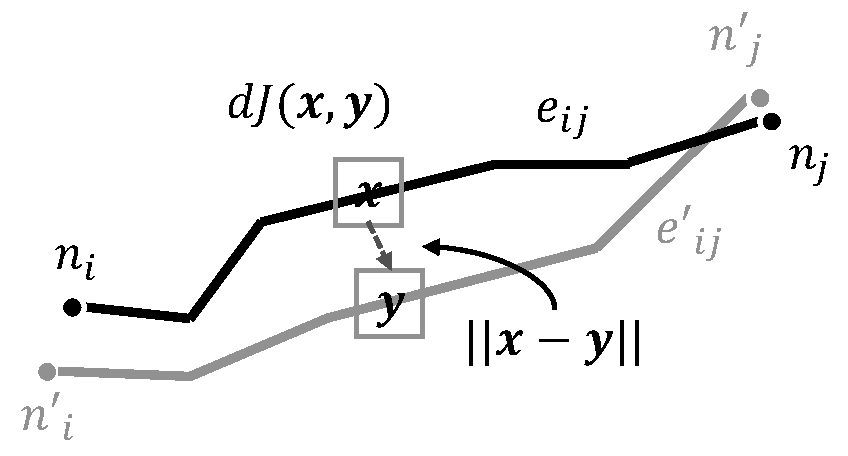}\\
 \vskip\baselineskip
  \end{center}
  \caption{Distance term and weight term for link $e_{ij}$}
 \label{transp}
\end{figure}

In the TG distribution, the distance term $||\boldsymbol{x}-\boldsymbol{y}||$ represents the geographic distance between road sections $\boldsymbol{x}$ and $\boldsymbol{y}$.
As shown in Figure \ref{transp}, the transported mass $dJ(\boldsymbol{x},\boldsymbol{y})$ is moved from position $\boldsymbol{x}=(p_A,q_A)$ in $D_A'$ to position $\boldsymbol{y}=(p_B,q_B)$ in $D_B'$.
Thus, the distance term has the dimension of length.

The weight term $dJ(\boldsymbol{x},\boldsymbol{y})$ represents the traffic-weighted quantity associated with link $e_{ij}$.
To enable continuous integration, this quantity must be divisible.
Let $w_{ij}$ denote the traffic-related weight of link $e_{ij}$.
Since traffic volume itself is not spatially accumulable, i.e., the sum of traffic volumes over subdivided segments does not equal the original link volume, it cannot be directly used as $w_{ij}$.
Instead, the total travel distance $q_{ij} \times C$ is used as the weight, which is spatially accumulable.
Accordingly, the dimension of the weight term is vehicle $\times$ length.

As a result, the unit of $TGW(A,B)$ becomes length$^2 \times$ vehicle.
A small TGW distance indicates that major links with high traffic volumes are geographically well aligned between two networks.
In contrast, large values arise when heavily used routes are significantly displaced or detoured due to positional or topological inconsistencies.

Moreover, the optimal transport formulation not only provides a scalar distance for global (TGW distance), but also describes how traffic mass is transported between locations.
This transport plan forms the basis of TG-OTM, which enables spatial interpretation of differences by revealing where large amounts of traffic-related mass are moved over long distances.
Therefore, the proposed method supports both global quantification of differences and local identification of their spatial patterns.

\subsubsection{Evaluating completeness}

Completeness evaluates the extent to which links and nodes are present in a road network dataset compared to a reference.
Existing metrics typically measure completeness as the presence or absence of geometric elements, without considering their impact on traffic states \citep{Haklay_2010, Hashemi_2017}.

In the proposed method, completeness is evaluated through its effect on traffic state distributions.
If a link is missing, no traffic can be assigned to its location, resulting in zero traffic weight at that position.
Missing links on major roads, which carry large traffic weights, lead to significant discrepancies in the TG distribution and consequently large TGW distances.

Moreover, missing nodes may alter traffic assignment results.
For example, centroid nodes used for trip generation may differ between datasets, potentially changing route choices and traffic flow patterns \citep{Qian_2012}.
Such changes further contribute to differences in traffic state distributions.

These effects can also be localized using TG-OTM, where missing high-traffic links appear as areas with large transported mass.

\subsubsection{Evaluating positional accuracy}

Positional accuracy evaluates the spatial accuracy of links and nodes in a road network dataset.
Existing approaches measure this aspect by the average geometric distance between matched links in two datasets \citep{Hashemi_2017}.

In the proposed method, positional differences directly correspond to the distance term $||\boldsymbol{x}-\boldsymbol{y}||$ in the Wasserstein distance.
Small positional differences result in small transport distances and thus a small TGW distance value, whereas large misalignments increase the transport cost.

Therefore, TGW distance reflects the overall positional consistency between two datasets.
In addition, TG-OTM reveals where such positional differences are concentrated, particularly when large traffic volumes are associated with misaligned links.

\subsubsection{Evaluating topological correctness}

Topological correctness evaluates the connectivity of links and nodes in a road network dataset.
Existing methods often assess topology based on node degrees or connectivity measures \citep{Haklay_2010, Hashemi_2017}.

In contrast, the proposed method evaluates topology through its impact on traffic flow patterns.
Topological errors, such as missing connections, can alter shortest paths and force traffic to take detours.
Such detours increase both the geographic distance to the original route without detour and total travel distance, leading to larger values in both the distance term $||\boldsymbol{x}-\boldsymbol{y}||$ and the weight term $dJ(\boldsymbol{x},\boldsymbol{y})$.
As a result, topological inconsistencies produce large TGW distances.

These effects are also observable in TG-OTM as concentrated transportations of transported mass between original and alternative routes, enabling the identification of locations where topological issues significantly affect traffic states.

\subsubsection{Advantages and disadvantages of the proposed method}

By transforming road network datasets into their TG distributions, the proposed method enables comparison based on traffic-weighted geographic characteristics, thereby evaluating the performance of road network datasets in transportation analysis at a macroscopic scale.
Rather than assessing individual quality aspects directly, it evaluates their combined effects on overall traffic state distributions.

In addition to the global comparison using TGW distance, the proposed method provides a local interpretation through TG-OTM.
While TGW distance quantifies the overall magnitude of differences, TG-OTM reveals where and how these differences occur by identifying spatial patterns of transported mass.
In particular, areas with large transported mass and long transport distance indicate locations where discrepancies in traffic states are significant, allowing the method to detect and localize potential data quality issues.

However, this macroscopic framework also has limitations.
When multiple quality issues (e.g., completeness, positional errors, and topological inconsistencies) coexist, their effects on traffic state distributions may overlap, making it difficult to attribute observed differences to a specific cause.

The proposed method is not intended to replace existing comparison approaches that focus on individual quality aspects, nor does it conflict with them.
Instead, it complements such methods by providing an integrated perspective on overall network performance, particularly from the viewpoint of traffic impacts.
Therefore, it can be used combined with existing methods for a more comprehensive evaluation of road network datasets.

\section{Case study}
In the case study, we validate if the proposed method could quantitatively measure global difference and find local differences of traffic states distributions between two road network datasets.
Datasets with different extractions and from different sources are used for the validation.


\subsection{Purpose}
The purpose of this case study is to validate the proposed method from both global and local perspectives.
Specifically, we aim to demonstrate that:
\begin{enumerate}
\item  TGW distance can be used as a indicator of global differences in traffic state distributions.
\item  Spatial patterns of TG-OTM can be used to detect local network problems.
\end{enumerate}

\subsection{Experiment settings}

Two experiments are designed to evaluate the two capabilities of the proposed method.
The first experiment evaluates global differences by analyzing how TGW distance changes under different levels and types of network extraction.
The second experiment evaluates local differences by analyzing the spatial patterns of TG-OTM between networks from different data sources.

In this case study, we focus on network extraction, a specific kind of network simplification that constructs a reduced network by retaining a subset of links from the original network, without aggregation or abbreviation.
We define a reference network as the baseline network against which other networks are compared, and a comparison target network as an extracted network whose traffic state distribution is evaluated relative to the reference.

\subsubsection{Datasets}

Two road network datasets are used in the experiments: OpenStreetMap (OSM) and the Digital Road Map (DRM).
OSM is a widely used open-source dataset of volunteered geographic information provided by the OpenStreetMap Foundation.
DRM is a surveying-based dataset developed from road network data provided by road administrators and is widely used as a standard road network dataset in Japan.
Although both datasets cover the same area, they include minor differences in link positions and road type classifications due to differences in data sources and mapping processes.

Both datasets organize roads into a class hierarchy, but use different naming.
The road hierarchy ordered as motorway $>$ trunk $>$ primary $>$ secondary $>$ tertiary $>$ residential in OSM, and expressway $>$ national road $>$ primary prefectural road $>$ prefectural road $>$ basic road $>$ narrow streets, correspondingly in DRM.

\subsubsection{Global difference}
This experiment evaluates whether the proposed method can quantify the magnitude of differences caused by network extraction.

\begin{figure}
  \begin{center}
    \includegraphics[width=17cm]{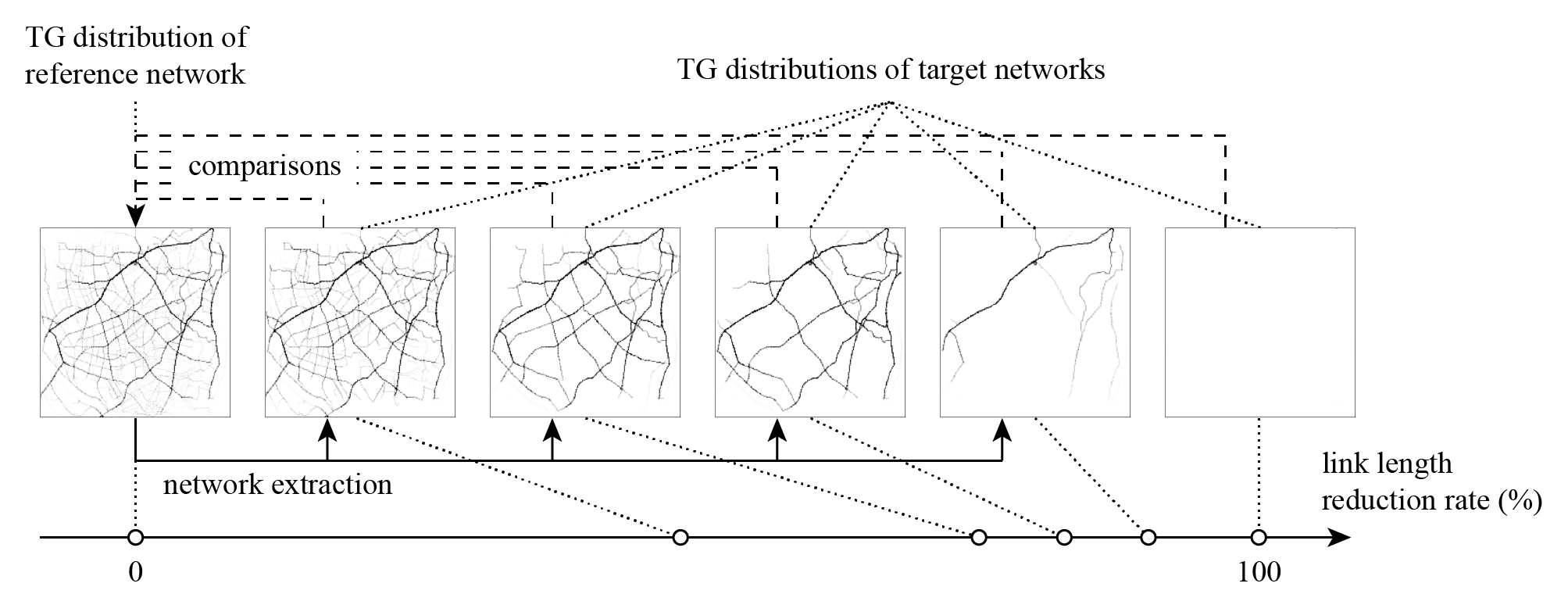}\\
 \vskip\baselineskip
  \end{center}
  \caption{Comparison for analyzing road networks with network extraction}
 \label{mtd}
\end{figure}

In both series, as shown in Figure \ref{mtd}, the full-detail OSM network is used as the reference network, and extracted networks are used as comparison targets.
The full-detail network refers to the original network derived from the dataset, in which all routable links are retained and routing is possible between any origin-destination pair.
Note that the TG distribution of this reference network does not represent true real-world traffic, but rather the traffic states simulated directly from the original dataset.
Based on this reference network, two series of road networks generated by different extraction methods are analyzed.
The expected changes in traffic state distributions can be qualitatively predicted for each series.

\begin{table}
 \caption{Comparison cases of road-type-based extraction (reference: full-detail OSM)}
 \label{tbl:cases-roadtype}
  \centering
  \begin{tabular}{llll}
 \hline
case ID & comparison target network & retained road classes & description \\
\hline
G1 & tertiary-level OSM  & motorway, trunk, primary, secondary, tertiary & finest extraction \\
G2 & secondary-level OSM & motorway, trunk, primary, secondary& medium extraction  \\
G3 & primary-level OSM   & motorway, trunk, primary & coarse extraction \\
G4 & trunk-level OSM     & motorway, trunk & coarsest extraction \\
\hline
\end{tabular}
\end{table}

First, a series of road-type-based extracted OSM networks (Table \ref{tbl:cases-roadtype}) is used to represent appropriate extraction, where traffic state distributions are expected to remain similar.
Following the road hierarchy, each extracted network is named by the lowest road class it retains, listed in order of increasing detail: \textit{trunk-level} (motorway and trunk), \textit{primary-level} (additionally primary), \textit{secondary-level} (additionally secondary), and \textit{tertiary-level} (additionally tertiary).
Coarser extractions retain only higher-class roads, while finer extractions also include lower-class roads, corresponding to different levels of network abstraction.

\begin{table}
 \caption{Comparison cases of random link-reduction extraction (reference: full-detail OSM)}
 \label{tbl:cases-random}
  \centering
  \begin{tabular}{lll}
 \hline
case ID & comparison target network & description \\
\hline
G5 & 20$\%$ link-reduced OSM & 20$\%$ of links (by count) randomly removed \\
G6 & 40$\%$ link-reduced OSM & 40$\%$ of links (by count) randomly removed \\
G7 & 60$\%$ link-reduced OSM & 60$\%$ of links (by count) randomly removed \\
G8 & 80$\%$ link-reduced OSM & 80$\%$ of links (by count) randomly removed \\
\hline
\end{tabular}
\end{table}

In contrast, a series of randomly link-reduced OSM networks (Table \ref{tbl:cases-random}) is used to represent network degradation, where traffic state distributions are expected to change significantly.
In this case, a $k\%$ \textit{link-reduced network} is constructed by randomly removing $k\%$ of the links by count, regardless of road class.
The resulting reduction in total link length differs from $k$ and is reported separately in the results.
This process often breaks network structure and introduces topological errors, even when the overall completeness appears high.

For each target network, the TGW distance to the reference network is calculated based on their TG distributions.
The relationship between TGW distance and the reduction rate of total link length is then analyzed.
Here, the reduction rate is defined as the proportion of total link length removed from the reference network.
TGW distance represents the difference in traffic state distributions, while the reduction rate represents the aggressiveness of network extraction.

\subsubsection{Local differences}
This experiment evaluates whether the proposed method can locate and explain differences that have a large impact on traffic state distributions.

Road networks from two different data sources, OSM and DRM, are compared to identify local differences.
In real datasets, multiple quality issues, such as missing links and positional errors, often exist at the same time.
Their effects on traffic states may overlap and are expected to be significant in certain locations.

\begin{table}
 \caption{Comparison cases between OSM and DRM at different network scales (cross-source comparison)}
 \label{tbl:cases-source}
  \centering
  \begin{tabular}{llll}
 \hline
case ID & network scale & OSM side (reference) & DRM side (comparison target) \\
\hline
L1 & full-detail        & full-detail OSM       & full-detail DRM       \\
L2 & secondary-level    & secondary-level OSM   & prefectural-road-level DRM   \\
L3 & trunk-level        & trunk-level OSM       & national-road-level DRM       \\
\hline
\end{tabular}
\end{table}

Three pairs at the full-detail, secondary-level, and trunk-level scales (Table \ref{tbl:cases-source}) are extracted based on the OSM road hierarchy.
Each pair consists of networks covering the same study area but derived from different data sources.

For each pair, the TG-OTM between their TG distributions is computed and visualized.
Areas with large transported mass and long transport distance indicate locations where differences in traffic state distributions are significant, suggesting potential data quality issues such as missing links or positional errors.

\subsection{Study area and parameter settings}

\subsubsection{Study area and datasets}

\begin{figure}
  \begin{center}
    \includegraphics[width=12cm]{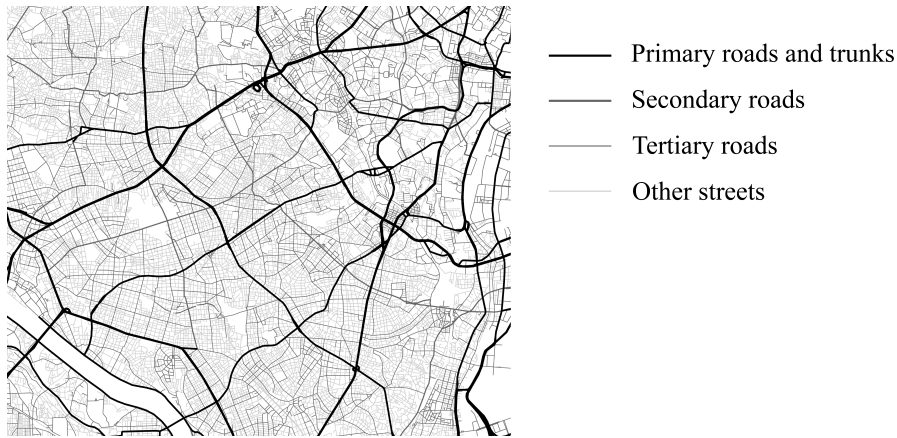}
  \end{center}
  \caption{OSM network of study area in southwest Tokyo, Japan}
  \label{area_osm}
\end{figure}

The case study uses road network datasets within a 10 km $\times$ 10 km study area in southwest Tokyo, Japan (Figure \ref{area_osm}).
The OSM data used in this study were obtained in 2024, and the DRM data correspond to the March 2024 release.

\subsubsection{Parameter settings}

Link lengths are directly obtained from the datasets.
Other traffic-related parameters, including limit speeds and link capacities, are assigned based on road type for both OSM and DRM networks.
To enable cross-source comparison, the corresponding OSM and DRM classes are treated as equivalent for parameter assignment, as shown in Table \ref{tbl8}.
Speed limit and capacity are assigned to each link from this table based on its road class.

\begin{table}
 \caption{Parameter settings of limit speed and capacity}
 \label{tbl8}
  \centering
  \begin{tabular}{llrr}
 \hline
road type (OSM) & road type (DRM) & limit speed (kph) & capacity (veh/hour) \\
\hline
motorway & expressway & 100 & 10,000 \\
trunk & national road & 60 & 7,500 \\
primary & primary prefectural road & 60 & 2,500 \\
secondary & prefectural road & 60 & 1,500 \\
tertiary & basic road & 40 & 500 \\
residential & narrow streets & 40 & 250 \\
\hline
\end{tabular}
\end{table}

Empirical parameters of BPR function (equation \ref{bpr}) are set as $\alpha=0.48, \beta=2.82$, following previous studies in Japan \citep{JSCE_2003}.

To ensure fair comparison, the same travel demand is used for all networks.
The OD demand is estimated from the full-detail OSM network and applied to all cases.

\begin{figure}
\begin{center}
\includegraphics[width=6cm]{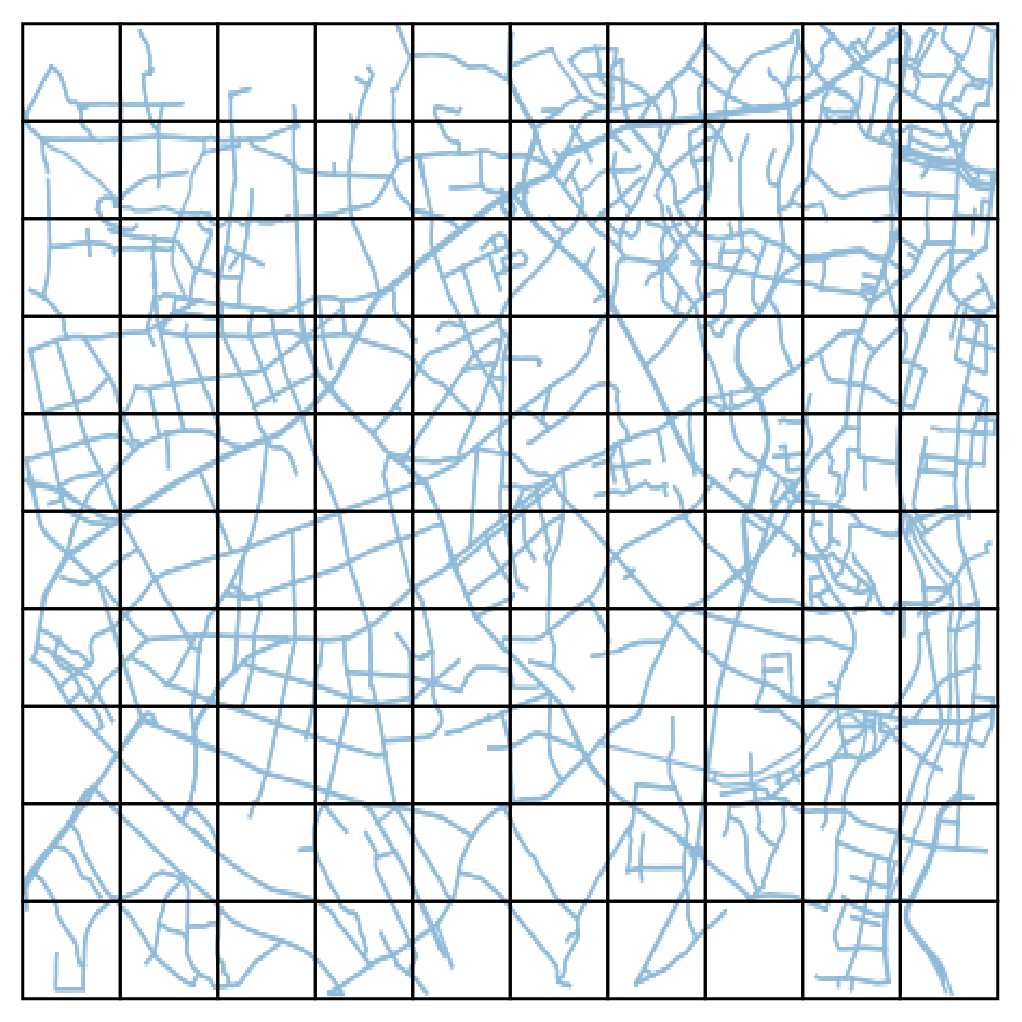}
\caption{Zones in the study area}
\label{zones}
\end{center}
\end{figure}

The study area is divided into 1 km $\times$ 1 km grids, resulting in 100 zones (Figure \ref{zones}).
A daily traffic condition is assumed.
For simplicity, all zones are assigned the same trip generation and attraction rates ($G = A = 1000$ veh/hour).
Additionally, zones connected to major external roads are assigned additional demand (5000 veh/hour per major road).

The OD flow $t_{ij}$ is calculated using the gravity model (Equation \ref{gravity}) with parameters $\alpha = 0.5$, $\beta = 0.5$, $\gamma = -0.5$, and $k = 0.1$.

For TG distribution discretization, the grid size $C$ is set to 50 m, resulting in a resolution of 200 $\times$ 200.
The penalty parameter $\lambda$ in the unbalanced Wasserstein distance is set to 0.05 veh$\times$km$^2$.
This value corresponds to the cost of moving traffic over a long distance and reflects the assumption that drivers do not make large detours (e.g., around 1 km) from their original routes.

\subsection{Results of evaluating global difference}
\begin{figure}
  \begin{center}
    \includegraphics[width=16cm]{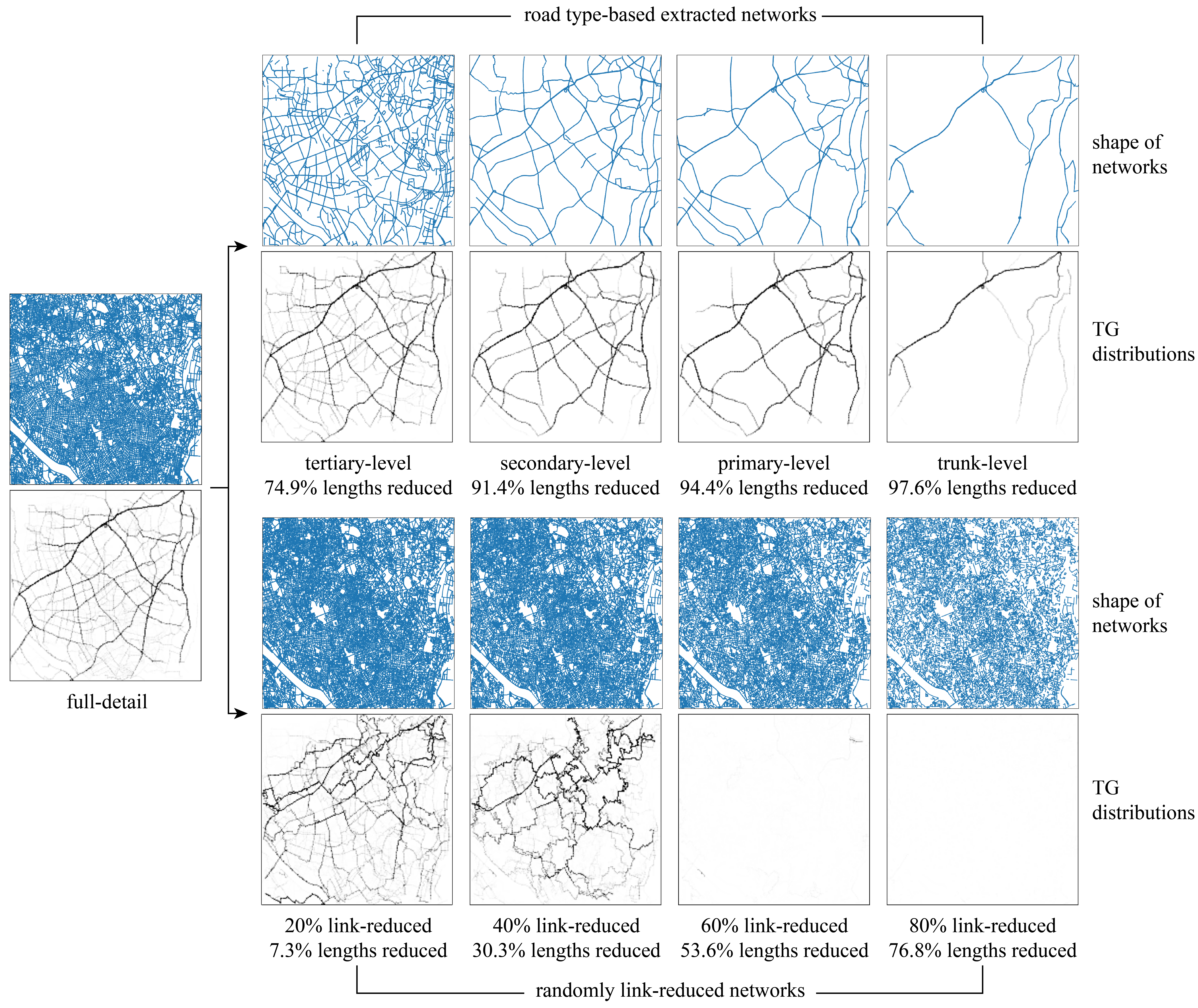}\\
 \vskip\baselineskip
  \end{center}
  \caption{Networks and TG distributions of different extracted OSM networks}
  \label{g1}
\end{figure}

Road networks and their corresponding TG distributions for both road-type-based extraction and random link-reduction extraction are shown in Figure \ref{g1}.

For road-type-based extraction, although a large number of links are removed, the overall shape of the TG distributions changes only slightly, and the main structure of major roads remains clear in all cases.
This indicates that traffic state distributions are largely preserved under this type of extraction.

In contrast, for randomly link-reduced networks, the TG distributions change significantly even with a smaller level of link reduction.
This reflects that random removal of links disrupts network structure and leads to large changes in traffic state distributions.

\subsubsection{TGW distance}

\begin{table}
 \caption{TGW distance for road-type-based extraction cases (reference: full-detail OSM)}
 \label{tbl:tgw-roadtype}
  \centering
  \begin{tabular}{llrr}
 \hline
case ID & comparison target network & link-length reduction vs.\ reference ($\%$) & TGW distance (veh$\times$km$^2$) \\
\hline
G1 & tertiary-level  & 74.9 &   95,266.0 \\
G2 & secondary-level & 91.4 &  342,929.9 \\
G3 & primary-level   & 94.4 &  595,740.7 \\
G4 & trunk-level     & 97.6 & 1,747,204.2 \\
\hline
\end{tabular}
\end{table}

\begin{table}
 \caption{TGW distance for random link-reduction extraction cases (reference: full-detail OSM)}
 \label{tbl:tgw-random}
  \centering
  \begin{tabular}{lrrr}
 \hline
case ID & link-count reduction $k$ ($\%$) & resulting link-length reduction ($\%$) & TGW distance (veh$\times$km$^2$) \\
\hline
G5 & 20 &  7.3 &  842,455.4 \\
G6 & 40 & 30.3 & 1,141,283.8 \\
G7 & 60 & 53.6 & 3,194,120.2 \\
G8 & 80 & 76.8 & 3,446,695.1 \\
\hline
\end{tabular}
\end{table}

\begin{figure}
  \begin{center}
    \includegraphics[width=14cm]{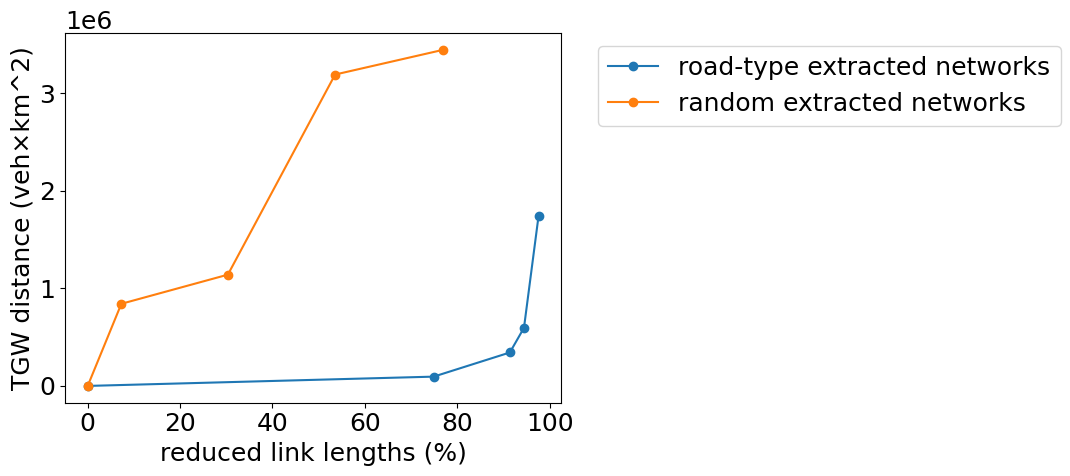}\\
 \vskip\baselineskip
  \end{center}
  \caption{Relationship between link length reduction rates and TGW distances of extracted OSM networks to the original OSM network}
  \label{f1}
\end{figure}

Table \ref{tbl:tgw-roadtype}, Table \ref{tbl:tgw-random}, and Figure \ref{f1} show the relationship between link length reduction rate and TGW distance from the reference network for different extracted networks.

For networks extracted by road type, TGW distances remain relatively small while high link reduction rates are achieved.
This indicates that traffic state distributions change only slightly.
In contrast, for randomly link-reduced networks, the TGW distance increases rapidly as the reduction rate increases.
This is because important links, including major roads, are removed randomly, which degrades network structure and causes large changes in traffic state distributions.

These results show that the proposed TGW distance can distinguish between appropriate extraction and harmful network degradation based on their impact on traffic states.
This confirms its effectiveness as a measure of global difference between road networks.

\subsubsection{TG-OTM}

\begin{figure}
  \begin{center}
    \includegraphics[width=13cm]{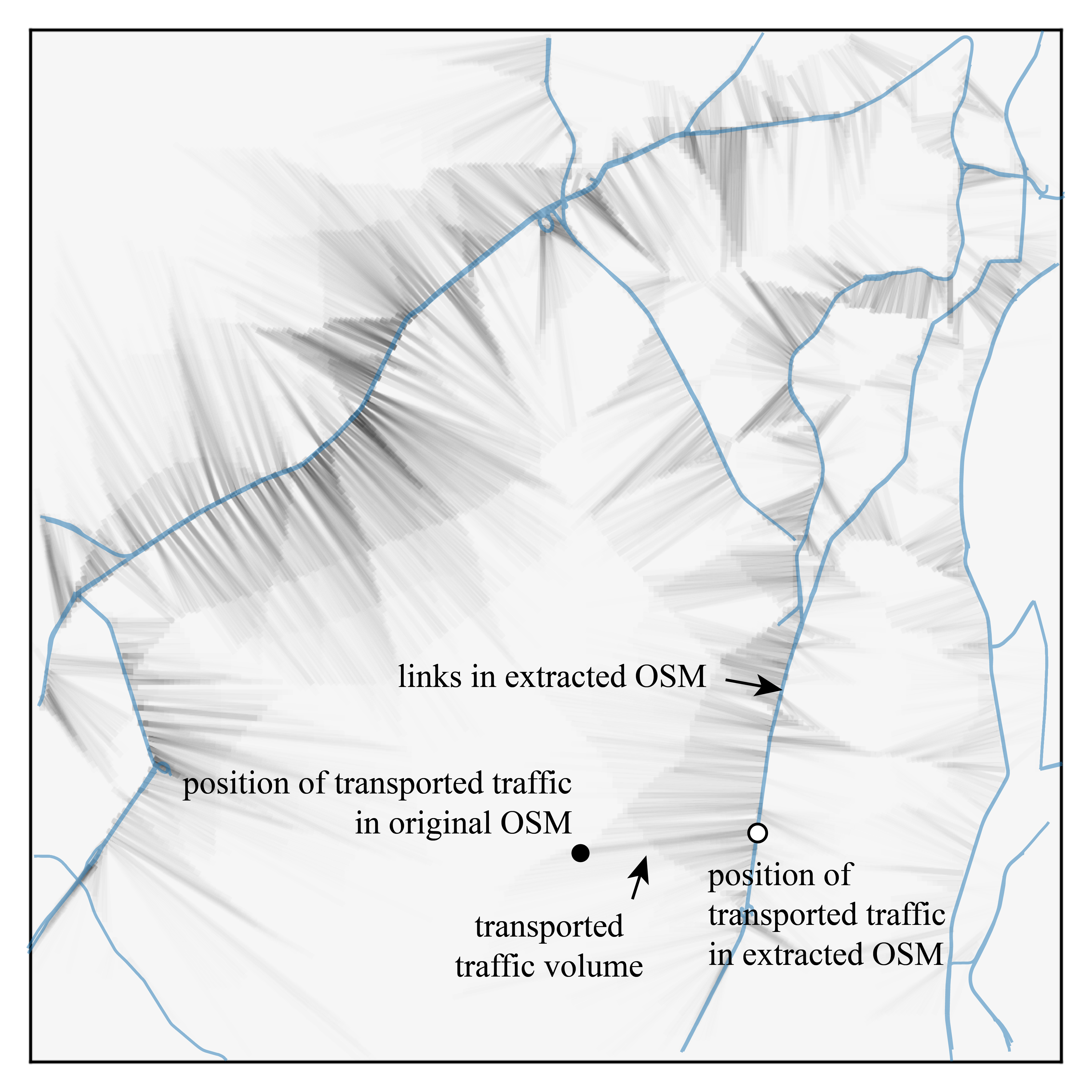}\\
 \vskip\baselineskip
  \end{center}
  \caption{TG-OTM of full-detail and trunk-level OSM networks}
  \label{f11}
\end{figure}

\begin{figure}
  \begin{center}
    \includegraphics[width=13cm]{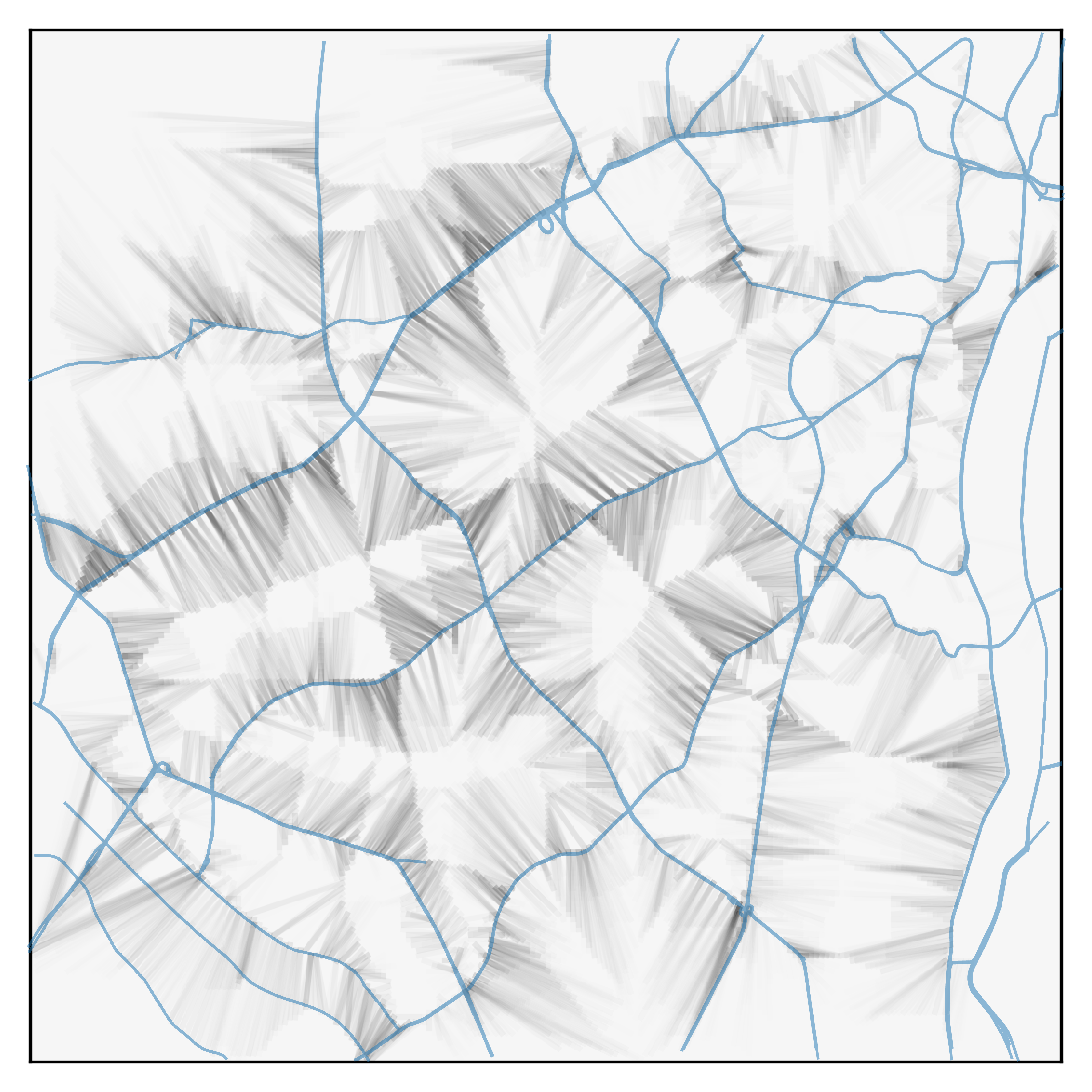}\\
 \vskip\baselineskip
  \end{center}
  \caption{TG-OTM of full-detail and primary-level OSM networks}
  \label{f12}
\end{figure}

\begin{figure}
  \begin{center}
    \includegraphics[width=13cm]{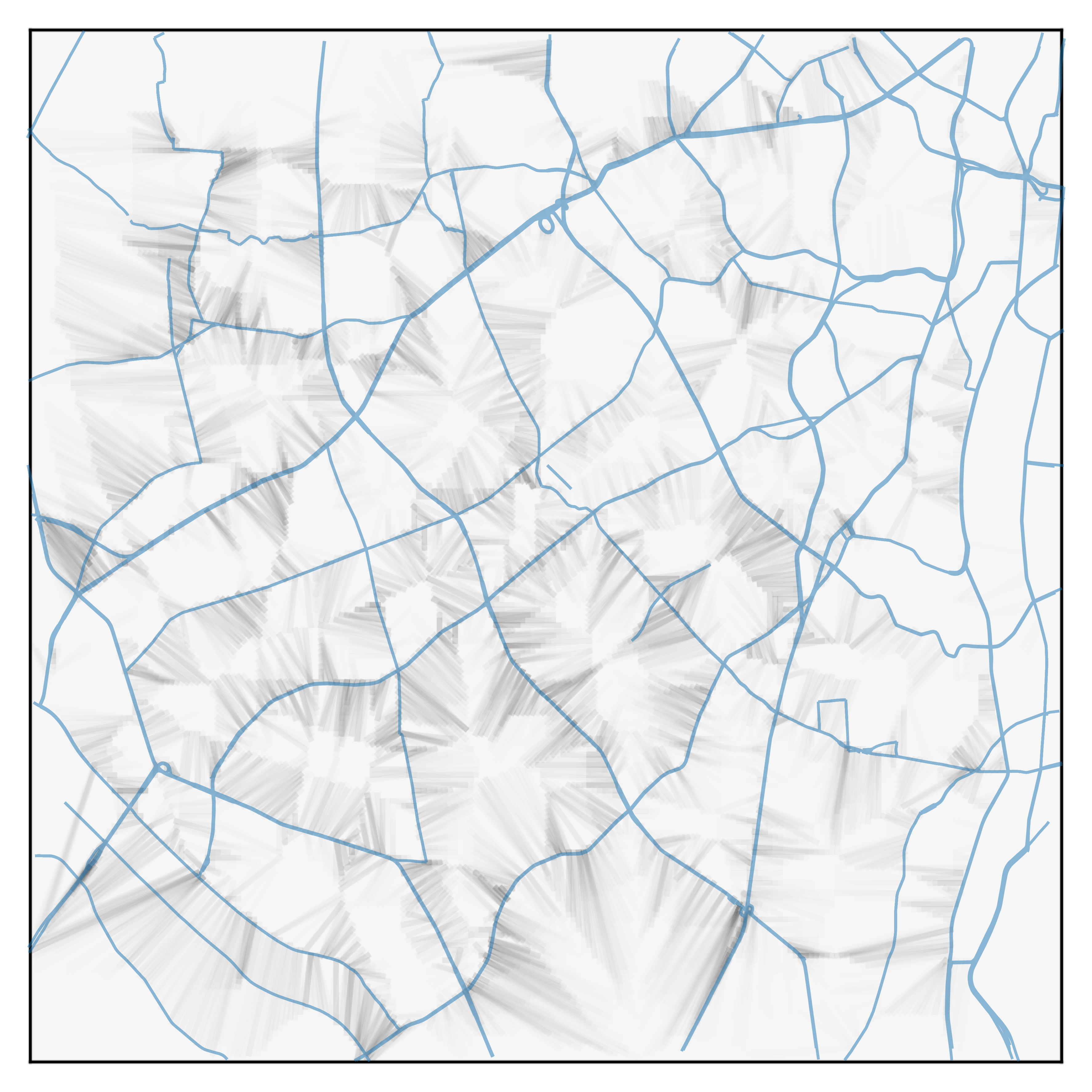}\\
 \vskip\baselineskip
  \end{center}
  \caption{TG-OTM of full-detail and secondary-level OSM networks}
  \label{f13}
\end{figure}

\begin{figure}
  \begin{center}
    \includegraphics[width=13cm]{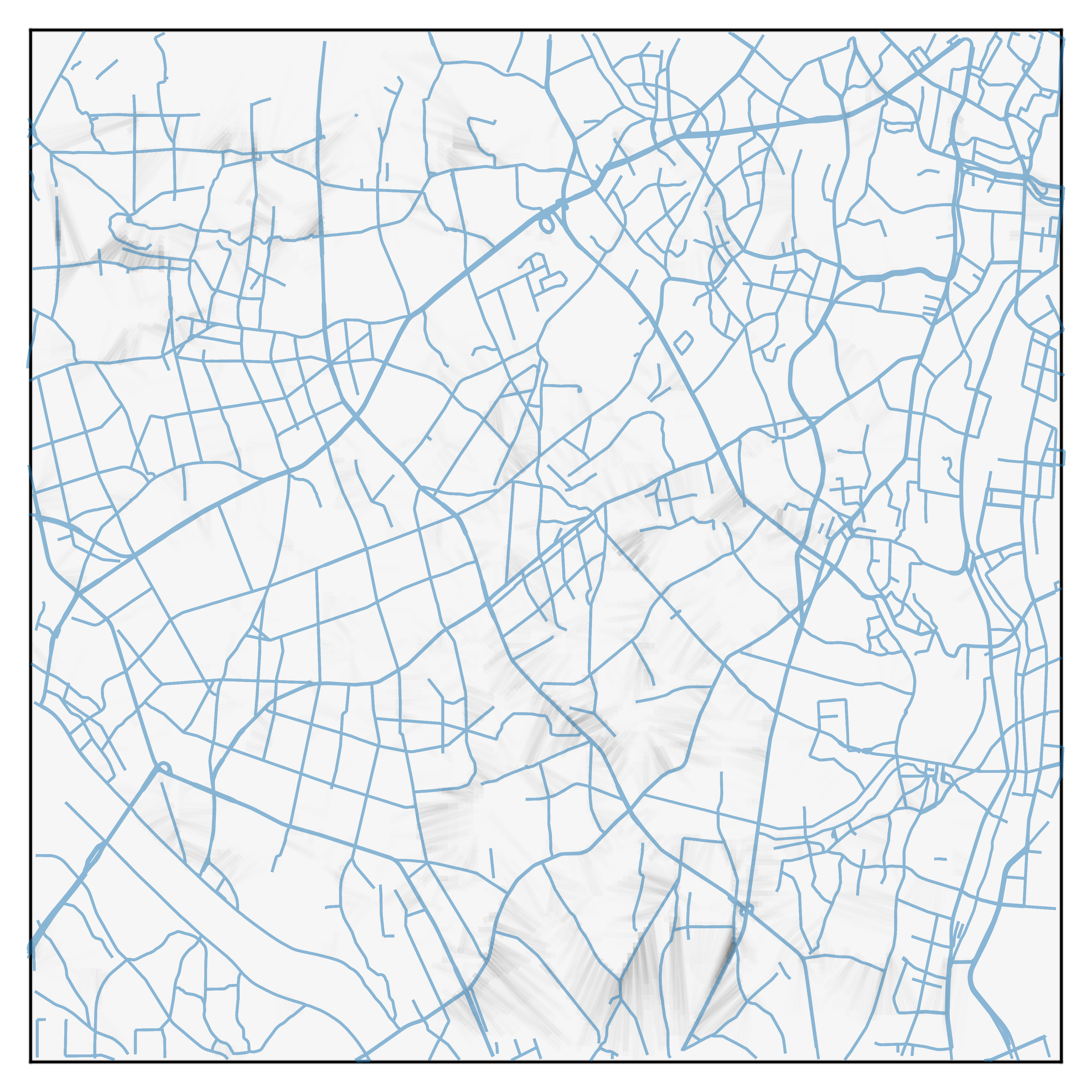}\\
 \vskip\baselineskip
  \end{center}
  \caption{TG-OTM of full-detail and tertiary-level OSM networks}
  \label{f14}
\end{figure}

Figures \ref{f11} to \ref{f14} show the TG-OTM between the reference network and the extracted networks.
Each line segment represents transported traffic mass, where the start point is a location in the reference network and the end point is the corresponding location in the extracted network.
Longer lines indicate longer transport distance, and darker color indicates larger transport mass.

The results show that traffic mass is mainly moved from areas of removed minor roads to nearby major roads.
This reflects a common pattern of traffic reassignment caused by network extraction.
For coarsely extracted networks (e.g., Figure \ref{f11}), large transport distance and mass are widely observed, which corresponds to large TGW distance values.
In contrast, for finely extracted networks (e.g., Figure \ref{f14}), there are less transport with shorter distances and smaller mass.
This indicates that the overall traffic distribution is preserved.

These results show that the TG-OTM can visually explain how and where traffic state distributions change between networks.
Together with the TGW distance, the proposed method can both quantify the magnitude of global differences and reveal their spatial patterns.

\subsection{Results of detecting local differences}

\subsubsection{TGW distance}

\begin{table}
 \caption{TGW distance between OSM and DRM networks at different network scales}
 \label{tbl:tgw-source}
  \centering
  \begin{tabular}{llr}
 \hline
case ID & network scale (extracted network) & TGW distance (veh$\times$km$^2$) \\
\hline
L1 & full-detail        & 143,397.6 \\
L2 & secondary-level    & 203,532.6 \\
L3 & trunk-level        & 105,695.4 \\
\hline
\end{tabular}
\end{table}

Table \ref{tbl:tgw-source} shows the TGW distances between the OSM and DRM networks at different network scales.
The values are relatively small at all scales, indicating that the overall traffic state distributions are similar between the two networks.
This result suggests that the proposed method captures global similarity while allowing local differences to be further analyzed.

\subsubsection{TG-OTM between full-detail networks}

\begin{figure}
  \begin{center}
    \includegraphics[width=13cm]{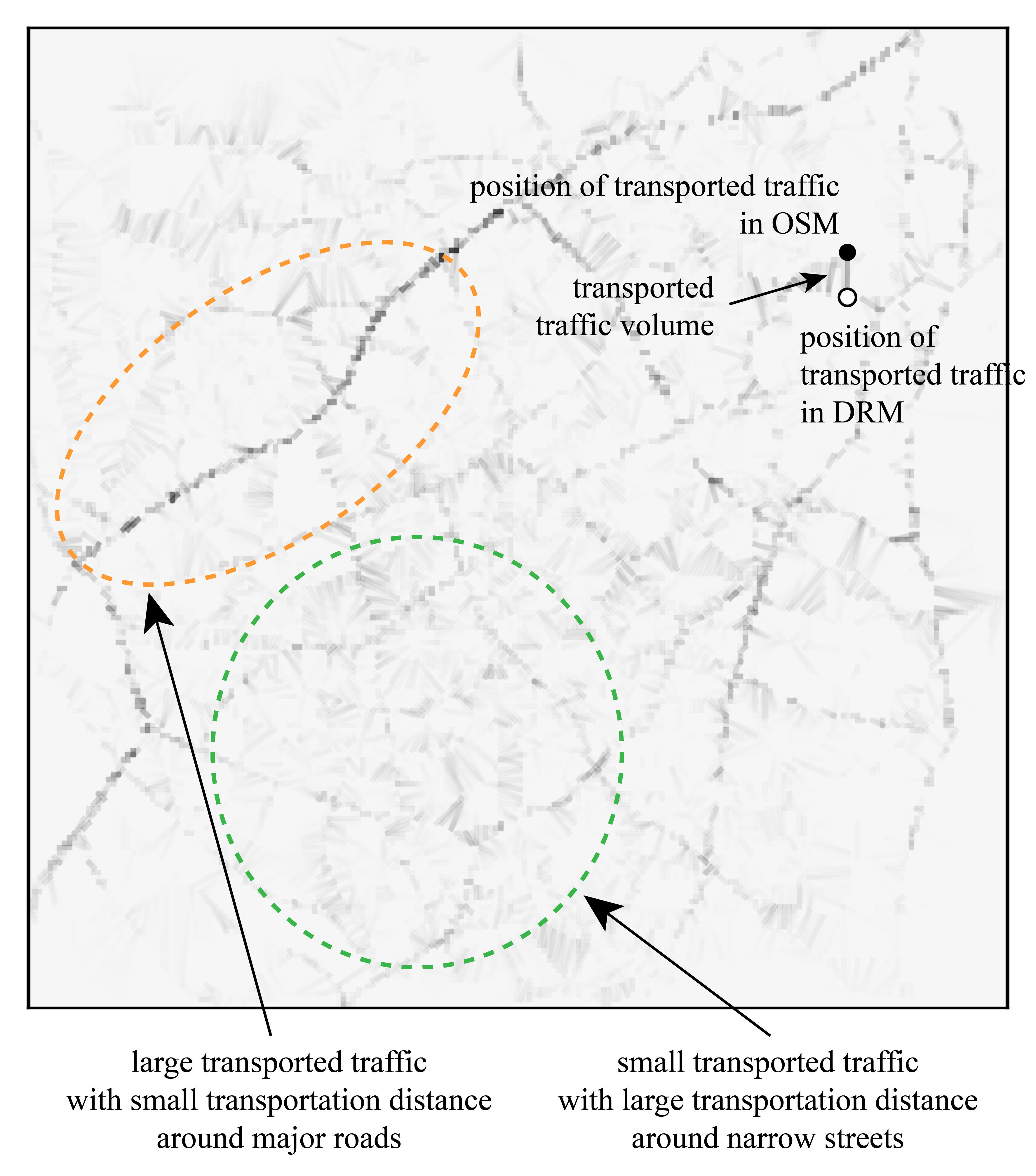}\\
 \vskip\baselineskip
  \end{center}
  \caption{TG-OTM of full-detail OSM and DRM networks}
  \label{f0}
\end{figure}

Figure \ref{f0} shows the spatial distribution of the TG-OTM between the full-detail networks of OSM and DRM.
Areas with both large transport distance and mass indicate large differences in traffic state distributions.

Because the full-detail networks contain large numbers of nodes and links, direct matching is difficult.
However, the proposed method compares the networks based on distribution of traffic states without node or link matching.
This demonstrates the applicability of the method to large and complex networks.

In the results, areas near major roads show large transport mass but short distances, reflecting small positional differences.
In contrast, local roads show longer distances due to missing or inconsistent links, but the transport mass is small because traffic is low.
As a result, only a few areas have both large mass and long distance, indicating that the overall traffic distributions are similar.
These results show that the proposed method can detect local differences while preserving global similarity, confirming its effectiveness for detailed network comparison.

\subsubsection{TG-OTM between secondary-level networks}

\begin{figure}
  \begin{center}
    \includegraphics[width=13cm]{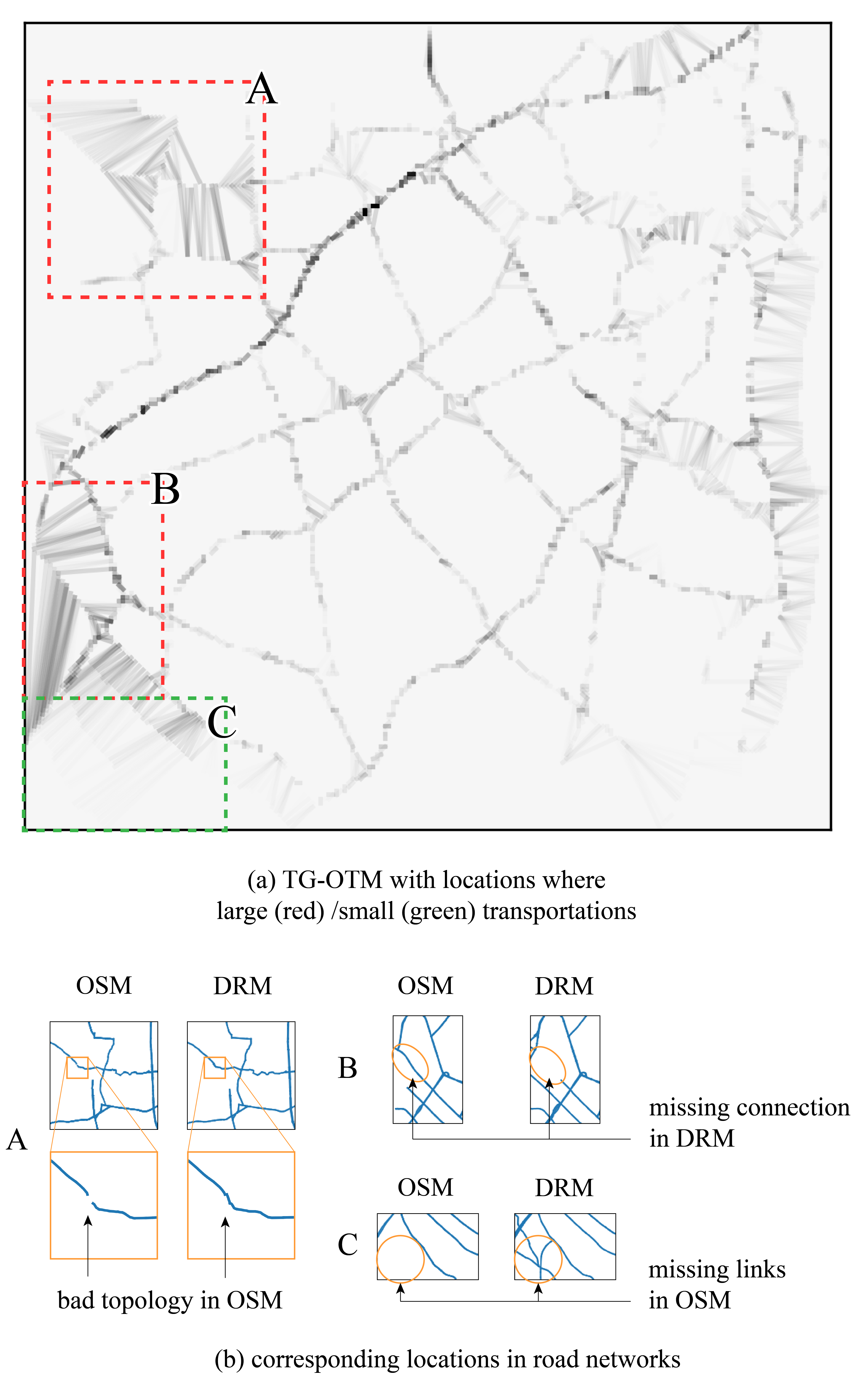}\\
 \vskip\baselineskip
  \end{center}
  \caption{TG-OTM of secondary-level OSM and DRM networks}
  \label{f3}
\end{figure}

Figure \ref{f3} shows the TG-OTM between the secondary-level networks of OSM and DRM.
We focus on three locations (A, B, and C) where differences in traffic states and network quality are observed.

At locations A and B, large transport mass and distance are observed, and clear network quality problems are found.
At location A, a short road segment is classified as a tertiary road in OSM, which breaks the connectivity between major roads during extraction, while DRM keeps the connection.
This inconsistency changes the traffic assignment results and appears as large transport in the TG-OTM.
Similarly, at location B, a connecting road is classified as a lower-class road in DRM and is removed.
This missing link causes detours and large changes in traffic states, which are also captured as large transport.

In contrast, at location C, missing links exist in OSM, but no large transport is observed.
This is because these links are not connected to major roads and have little impact on traffic assignment results.

These results show that the proposed method identifies network problems based on their impact on assigned traffic states.
It selectively detects differences that are important for traffic analysis.

\subsubsection{TG-OTM between trunk-level networks}

\begin{figure}
  \begin{center}
    \includegraphics[width=13cm]{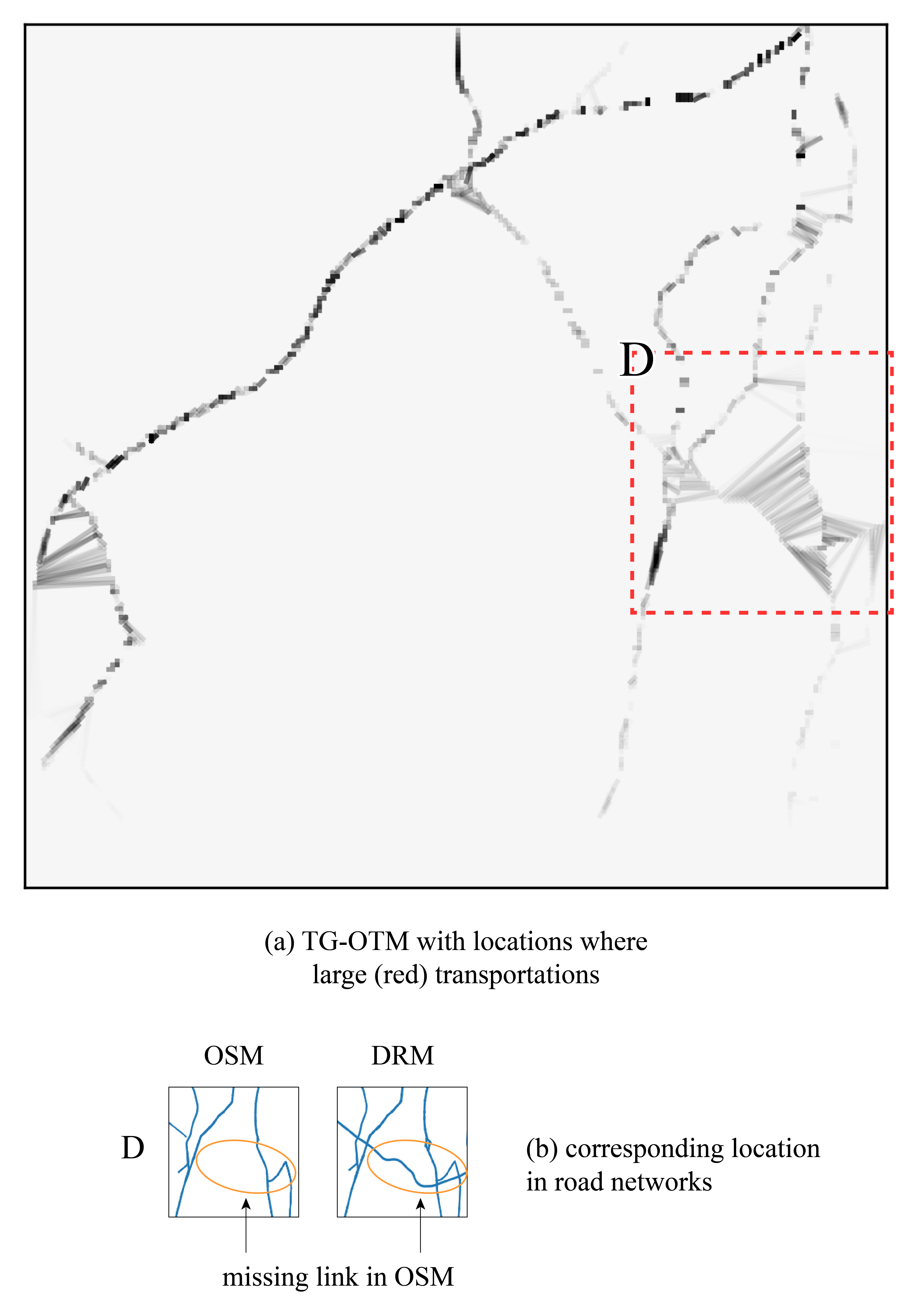}\\
 \vskip\baselineskip
  \end{center}
  \caption{TG-OTM of trunk-level OSM and DRM networks}
  \label{f4}
\end{figure}

Figure \ref{f4} shows the TG-OTM between the trunk-level networks.
At location D, both transport mass and distance are large, indicating a significant structural difference.

At this location, part of a highway is missing in OSM because its endpoints lie outside the study area and are removed during extraction.
In DRM, the link is kept because its endpoints lie on the boundary.
As a result, traffic that should pass through the highway is not reproduced in OSM and is reassigned to nearby roads, changing the traffic distribution.

The transport in TG-OTM does not always directly represent actual route changes or traffic loss due to the limitation of traffic assignment methods.
However, areas with large changes in traffic states still show large transport.
This demonstrates that the method correctly captures the magnitude of the impact of network differences.

\subsubsection{Summary}

The proposed method compares road networks based on traffic state distributions without matching nodes or links.
Areas with both large transport distance and mass correspond to network quality problems that strongly affect traffic states.
In contrast, differences with small impact do not appear as large transport.

These results show that the method can selectively detect important differences for traffic analysis.
Therefore, it is effective for evaluating network quality and identifying locations with problems that should be fixed with high priority.

\subsection{Summary}

The case study validates the proposed method for comparing road network datasets based on traffic state distributions.
The method is evaluated from both global and local perspectives using TGW distance and TG-OTM.

From the global perspective, the results show that TGW distance can effectively measure overall differences between networks.
It remains small for appropriately extracted networks where distribution of traffic states preserves, and increases significantly when network structure is degraded.
This confirms that TGW distance captures meaningful changes in traffic state distributions.

From the local perspective, the results show that TG-OTM can identify where important differences occurs.
Areas with both large transport mass and long transport distance correspond to network quality problems, such as missing links or incorrect topology, that have a strong impact on traffic states.
In contrast, differences with little impact on traffic do not appear as large transport.

These results demonstrate that the proposed method can both quantify the magnitude of global differences and identify their local causes.
Therefore, it is effective for evaluating the quality of road network data and for detecting locations that should be prioritized for correction.

\section{Conclusions}

In this study, we proposed a method for comparing road network datasets based on traffic state distributions at a macroscopic level.
By representing road networks as traffic-weighted geographic (TG) distributions, the method enables comparison through the traffic-weighted geographic Wasserstein (TGW) distance.
In addition, the traffic-weighted geographic optimal transport matrix (TG-OTM) shows the spatial transport of traffic mass between TG distributions.

The case study demonstrates that the proposed method can evaluate differences between networks from both global and local perspectives.
TGW distance provides a quantitative measure of overall differences in traffic state distributions, while TG-OTM reveals where these differences occur and how they are related to network structure.
The results show that the method can evaluate and detect meaningful differences that affect traffic from those with limited impact.

An important advantage of the method is that it does not require explicit matching of nodes or links between datasets, nor additional observed traffic data.
This makes it suitable for practical comparison and evaluation of road network datasets from different sources.

These findings suggest that the proposed method is effective for assessing the quality of road network data and for identifying areas that require improvement.

A future direction of this study will focus on extending the definition of traffic weights $w_{ij}$ to represent a wider range of traffic characteristics.
In addition to traffic volume, factors such as speed and density should be incorporated to better capture the complexity of traffic conditions.
This extension will improve the applicability of the method to various problems in transportation and urban network analysis.

\subsection*{\normalsize\bfseries Conflict of Interest Statement}\noindent
The authors declared no potential conflicts of interest with respect to the research, authorship, and/or publication of this article.

\subsection*{\normalsize\bfseries Acknowledgments}\noindent
This work was supported by the JST SPRING, Japan Grant Number JPMJSP2106.
The authors would like to thank the Japan Digital Road Map Association for providing the Digital Road Map (DRM) dataset used in this study.

\subsection*{\normalsize\bfseries Data availability statement}\noindent
The OpenStreetMap data used in this study are openly available from the OpenStreetMap project.
The Digital Road Map data used in this study cannot be made openly available because redistribution of the data is restricted.

\bibliographystyle{agsm}
\bibliography{ronbun-j}

\end{document}